\documentclass[12pt]{article}

\usepackage{epsf}
\renewcommand{\theequation}{\arabic{section}.\arabic{equation}}
\renewcommand{\thefootnote}{\fnsymbol{footnote}}

\newcommand{\Kd}{$K^0$--$\bar{K^0}$}
\newcommand{\Bd}{$B^0_d$--$\bar{B^0_d}$}
\newcommand{\Bs}{$B^0_s$--$\bar{B^0_s}$}
\newcommand{\model}{SU(2)$_L\times$SU(2)$_R\times$U(1)}
\newcommand{\ee}{$\epsilon'/\epsilon$}
\newcommand{\goldd}{$B^0_d\to J/\psi K^0_S$}

\newcommand{\vl}[2]{(V_L)_{#1#2}}
\newcommand{\vlc}[2]{(V_L^*)_{#1#2}}

\begin{document}

%%%%%%%%%%% Titlepage

\begin{titlepage}
\begin{flushright}
\begin{tabular}{l}
CERN--TH/99--297\\
ULB--TH--20\\
hep-ph/9910211
\end{tabular}
\end{flushright}
\vskip0.5cm
\begin{center}
  {\Large \bf 
Anatomy of 
Mixing-Induced CP Asymmetries
in Left-Right-Symmetric Models with Spontaneous CP Violation
  \\}

\vspace{1cm}
{\sc Patricia~Ball}${}^{1,}$\footnote{E-mail: Patricia.Ball@cern.ch},
{\sc J.-M.\ Fr\`{e}re}${}^{2,}$\footnote{E-mail: frere@ulb.ac.be} and
{\sc J. Matias}${}^{1,}$\footnote{E-mail: Joaquim.Matias@cern.ch}
\\[0.5cm]
\vspace*{0.1cm} ${}^1${\it CERN--TH, CH--1211 Geneva 23, Switzerland
}\\[0.3cm]
\vspace*{0.1cm} ${}^2${\it Service de Physique Th\'{e}orique, CP 225,
  Universit\'{e} Libre de Bruxelles, B--1050 Brussels, Belgium}\\[1.3cm]

%{\em Version of \today}

  \vfill

  {\large\bf Abstract\\[10pt]} \parbox[t]{\textwidth}{ 
We investigate the pattern of CP violation in \Kd, \Bd\ and \Bs\ mixing
in a symmetrical \model\ model with spontaneous CP violation. We calculate 
 the phases of the left and right quark mixing matrices beyond the
 small phase approximation and perform a careful analysis of all
relevant restrictions on the model's parameters from $\Delta m_K$,
$\Delta m_B$, $\epsilon$, \ee\  and the CP asymmetry in
\goldd. We find that, with current experimental data, the mass of the
right-handed charged gauge boson, $M_2$, is restricted to be in the
range 2.75 to 13$\,$TeV and the mass of the
flavour-changing neutral Higgs boson, $M_H$, in 10.2 to 14.6$\,$TeV. 
This means in
particular that the decoupling limit $M_2,M_H\to\infty$ is already
excluded by experiment. We also find that the model favours opposite
signs of $\epsilon$ and $\sin 2\beta$ and is excluded if $\sin2\beta > 0.1$.
}
  \vskip1cm 
{\em  submitted to Nuclear Physics B}\\[1cm]
\end{center}
\end{titlepage}

\setcounter{footnote}{0}
\renewcommand{\thefootnote}{\arabic{footnote}}

\section{Introduction}

In this paper we investigate in numerical detail an attractive 
extension
of the standard model (SM), the Spontaneously Broken Left-Right model
(SB--LR). The model turns out to be very constrained since, despite 
the larger number of bosons, symmetries strongly limit the new Yukawa
couplings. 

We show that a large fraction of parameter space is already excluded
by conservative bounds arising from the $B_d$ and $B_s$ mass differences,
the $\epsilon$ parameter for CP violation in the K system, and the sign 
of $\epsilon'$. Even if current theoretical uncertainties persist in the
K system, the expected
experimental progress in B physics will soon bring conclusive tests of
the
model.
 
It is well known that CP is a natural symmetry of pure gauge theories 
with massless fermions.\footnote{This stems from the unitary nature of
the
groups, or
equivalently the fact that gauge couplings are real. Anomalies can
bring in T or 
CP violation, but such "strong CP violation" is only defined with
respect to
the determinant of fermion masses.}
As a result, CP violation
actually 
probes the least known sector of unified theories, namely the scalar and
Yukawa couplings.
With the current development of dedicated accelerators to probe CP
violation in
the B system, it is important to study possible departures from the
"Standard Model"
based on the group SU$(2)_L \times $U(1). 
Models based on the group \model, and
more specifically
those exhibiting spontaneous CP violation, offer the advantage of a
well-defined,
and actually quite constraining context, largely testable
experimentally, while
presenting a structure significantly different from the Standard Model. 
Before going into any details, let us stress already that the
"left-handed" 
nature of the charged couplings in the SM, together with the absence of
CP violation in the neutral channels, is extremely constraining: for instance
electric dipole moments, intrinsically a LR transition, are strongly 
suppressed.
In the same line, the scalar potential for the SM seems incompatible
with
a first order electroweak transition, thereby hampering low-temperature 
baryogenesis.

While very close in many aspects to the SM, and for this reason a 
natural extension, "LR" models significantly depart from it 
and provide a rich structure both for laboratory CP violation and
baryogenesis
\cite{houart}, but also possibly in  the leptonic sector,
cf.~\cite{liu}; the
last two aspects 
will not be discussed in this paper.

By LR model we understand in general a description of electroweak interactions
based on the gauge group \model.
While such a group structure suggests low-energy parity restoration,
a necessary condition for this is the equality of gauge
couplings,
$g_L=g_R$.
This is not necessarily requested, and could actually prove a
difficulty,
notably in a cosmological approach: the persistence of an exact discrete
symmetry to low-energy can lead to the formation of domains
corresponding
to different orientations of the breaking, and consequently to
difficulties
with the walls-energy.
Some grand unified models, where   P, but \emph{not} \model\/
is broken at a very
high scale, lead to a low-energy structure where
$g_L \neq g_R$.
While in this paper we will for simplicity set $g_L=g_R$, the results
are easily adapted
to the more general case, as, due to the high mass of $W_R$, the
combination in
use is generally $g_R^2 / M_R^2$, or, for the mixing terms,
$g_L\cdot g_R \cdot \sin\zeta$, Ref.~\cite{langacker}, where $\zeta$ is
 the mixing-angle between L and R bosons.

Some symmetries are however needed in order to constrain the scalar and
Yukawa
sectors of the theory. We thus request P as a symmetry of the Lagrangian
(possibly, as stated above, in a weaker
form where the interchange of fermions $f_L \rightleftharpoons f_R$ is
accompanied
by $g_L \rightleftharpoons g_R$). Even the definition of this symmetry
is not
without ambiguities: indeed, with in general
non-diagonal mass matrices, the L and R partners are not uniquely
defined, namely,
a rotation $U$ in flavour space can be allowed for, namely $f_L
\rightleftharpoons U f_R$.
In order to restrict the model further, we implement the attractive
feature of
spontaneous CP violation.
This means that the Lagrangian must be symmetrical under CP (or
CP generalized to include $g_L \rightleftharpoons g_R$), which
is later broken by "misalligned" phases of the vacuum expectation
values.
Under these hypotheses, Ecker and Grimus have shown in Ref.~\cite{EGNPB}
that,
except for an
exceptional case (which will not be considered here), the Yukawa
couplings can be
parametrized in terms of two real symmetrical matrices.
As a result, all phases
of the model can be related to a unique phase affecting vacuum
expectation
values (noted $\alpha$ below) and calculated exactly. This point
is important, as it relates baryogenesis, and in particular the
sign of the matter-antimatter asymmetry, to low-energy CP violation  
\cite{houart}.
In practice, for the present analysis, four parameters are added to the
Standard 
Model, but in
counterpart, its single CP phase is now predicted.

The shorthand SB--LR will refer from now on to this "Spontaneously
Broken Left-Right model".
An important result of our analysis is that the SB--LR is in some sectors 
{\it more restrictive} 
than the SM itself. Indeed, while the SM is a subset of LR obtained by 
sending the
R sector masses to infinity, a similar procedure applied to the SB--LR
yields additional constraints since the CKM phase $\delta$ is no
longer independent, but predicted within the model. 
For instance, we find that the CP violating 
phase in the resulting  SM is too small, $|\delta|<0.25$ or 
$|\delta-\pi|<0.25$,  whereas the global fit
  of \cite{fit} yields $\delta= 1.0 \pm 0.2$. 
 Hence the SM limit
of the SB--LR is  inconsistent by 3.5$\,\sigma$
with current experiments. 
This has the important consequence that the SB--LR is actually testable,
and distinct from the SM: experimental bounds cannot be 
indefinitely evaded
by simply sending the R sector to infinite masses: scalars and vectors in the
range (2--20) TeV are definitely needed.

Experimental constraints on the SB--LR, mainly from the K system, have
been thoroughly investigated in \cite{langacker}. Since then, many SM
parameters, in particular the CKM angles and the top quark mass, have
been measured much more accurately, and the perspective of finding
non-standard CP violation in the B system at the B factories,
the Tevatron or the LHC has prompted a number of new analyses of the
SB--LR, for instance \cite{prades,bloedsinn}, which, however, all use
on a certain
approximation for calculating the phases of the left and right CKM
matrices. We thus feel that a comprehensive analysis of the 
constraints from measured CP conserving and violating observables in both
the K and the B system is timely, which in particular uses exact
expressions for the CKM
phases. The main new results of our analysis
can be summarized as follows:
\begin{itemize}
\item the small phase approximation fails for CKM matrix elements
  involving the 3rd generation;
\item the role of the Higgs bosons, neglected in most analyses, is crucial;
\item the decoupling limit of the model, $M_2,M_H\to\infty$, is
  experimentally excluded, which implies {\em upper bounds} on $M_2$
  and $M_H$;
\item the SB--LR favours opposite signs of the CP violating observables
  ${\rm Re}\,\epsilon$ and $a_{\rm CP}(B\to J/\psi K_S)$, which are both
  expected to be positive in the SM; hence, the model cannot accomodate both
  the experimentally measured $\epsilon$ and the SM expectation
  $a^{\rm SM}_{\rm CP}(B\to J/\psi K_S)\approx 0.75$ and is
  excluded if $a_{\rm CP}$ will be measured to be larger than 0.1.
\end{itemize}

The paper is organized as follows: in Sec.~2 we define the model
underlying our calculations. 
In Sec.~3 we calculate the phases of the
left and right quark mixing matrix. In Sec.~4 we formulate strategies
for 
measuring and/or constraining the SB--LR parameters. In Sec.~5 we
calculate 
B mixing
in the SB--LR and incorporate constraints from the measurement of the CP
asymmetry in $B^0\to J/\psi K_S$. In Sec.~6 we discuss constraints
on the model from K physics. In Sec.~7 we combine the constraints from both
K and B observables. In Sec.~8, finally, we summarize and conclude.

\section{Definition of the Model}
\setcounter{equation}{0}

We begin with a reminder, namely how   the extended gauge group \model\/
cascades down
to the unbroken electromagnectic subgroup U(1)$_{\rm em}$ through the
following simple symmetry-breaking pattern:
\addtolength{\arraycolsep}{-2pt}
$$
\underbrace{\begin{array}[t]{ccc} SU(2)_L & \times &
    SU(2)_R\\ T_L^i && T_R^i\\ g_L && g_R\end{array}} \times 
\begin{array}[t]{c} U(1)\\ S\\ g_1\end{array}$$
$$\makebox[1.5cm][c]{ }
\underbrace{\begin{array}{c@{\:\:}cc} SU(2)_L & \times & U(1)_Y\\ T_L^i
&&
    Y/2 = T_R^3 + S\\ g_L && g'\end{array}}
$$
$$\makebox[2cm][c]{ }
\begin{array}{c}U(1)_{\rm em}\\ Q = T_L^3 + Y/2\\e\end{array}
$$
\addtolength{\arraycolsep}{2pt}
Listed underneath each subgroup factor is our nomenclature convention
for their associated generators and coupling constants.

We next specify the quark and scalar content of the model. The quarks
transform under the unbroken gauge group as
\begin{equation}
q_{Li} = \left( \begin{array}{c} U_i\\D_i\\ \end{array}\right)_L \sim
(2,1,1/6),
\qquad q_{Ri} = \left( \begin{array}{c} U_i\\D_i\\ \end{array}\right)_R
\sim
(1,2,1/6),
\end{equation}
where $i$ is a generation index. 
The generation of quark masses in the \model\/ model 
requires at least one scalar
bidoublet $\Phi$, i.e.\ a doublet under both $SU(2)$, corresponding to
two standard
doublets:
\renewcommand{\arraystretch}{1.4}
$$
\Phi = \left( \begin{array}[c]{c@{\:\:}c} \phi_0^1 & \phi_1^+\\ \phi_2^-
&
    \phi_0^2 \end{array} \right)\sim (2,\overline{2},0).
$$
\renewcommand{\arraystretch}{1}
As usual, the quarks are given  masses by a spontaneous
breakdown of the symmetry such that $\Phi$ acquires the VEV
\renewcommand{\arraystretch}{1.2}
\begin{equation}
\langle \Phi \rangle = {1 \over{\sqrt{2}}}
\left(\begin{array}{c@{\:\:}c} v & 0\\ 0 &
w\end{array}\right).
\end{equation}
\renewcommand{\arraystretch}{1}
The quark mass matrices read
\begin{equation}\label{eq:MM}
M^{(u)} = {v\over{\sqrt{2}}}\Gamma + {w^*\over{\sqrt{2}}} \Delta, \quad
M^{(d)} = 
{w\over{\sqrt{2}}} \Gamma + {v^*\over{\sqrt{2}}} \Delta.
\end{equation}
In general, both $v$ and $w$ are complex, which is the source of
spontaneous CP
violation.
The phases can of course be redefined by a gauge rotation of the L or R
fields, and,
in particular, the phase \emph{difference} of $v$ and $w$ can be rotated
away.\footnote{This rotation was performed in \cite{JMF}.}
Here we follow the 
notations of \cite{EGNPB} and define \emph{two} phase combinations:
\begin{equation}
\left|\frac{w}{v}\right| = r, \qquad \arg(vw) = \alpha,  \qquad
\arg(-vw^*) = 
\lambda.\label{eq:xyz}
\end{equation}
Although the relevant diagrams are the same with an enlarged Higgs
sector, only in the minimal case are the couplings of scalar fields to
quarks determined by masses and mixing angles only. In this case,
neglecting for the moment the contributions from the triplets, which
do not couple directly to the quarks, 
there is a flavour-conserving neutral scalar field $\Phi_1$, the
analogue of the SM Higgs, and a single charged scalar field
$\Phi^{\pm}$ as well as two neutral scalar fields $\Phi_2$, $\Phi_3$
with flavour-changing couplings to quarks.

Further fields are needed to achieve the complete 
breakdown\footnote{Spontaneous breaking of CP without undue fine
  tuning may 
require
the introduction of further fields, for instance singlets uncoupled to the
fermions \cite{brancolavoura}. We will not discuss here the details of the
scalar Lagrangian, which is not critical for work.} from
\model\/ to U(1)$_{em}$; the simplest choices respecting LR symmetry
are either two doublets
(one L and one R), or two triplets. This latter
choice is
usually preferred when dealing with the leptonic sector, since the
quantum numbers
of these triplets allow for the generation of heavy neutrino Majorana
masses 
directly related to LR symmetry breaking; $b$ decays, however, do not
sensitively depend on this precise structure of the scalar
sector. The triplets are
$$
\chi_L = \left( \begin{array}{c} \chi_L^{++}\\ \chi_L^+\\ \chi_L^0
\end{array}
\right) \sim (3,1,2), \qquad \chi_R = \left( \begin{array}{c}
  \chi_R^{++} \\ \chi_R^+\\ \chi_R^0\end{array} \right) \sim (1,3,2)
$$
and acquire the VEVs
$$
\langle \chi_{L,R}\rangle = {1 \over{\sqrt{2}}}\left(
\begin{array}{c}0\\ 0\\v_{L,R} \end{array}
\right).
$$
In the remainder of this paper we shall assume $|v_L|^2 \ll |v|^2+|w|^2 \ll
|v_R|^2$.

The spontaneous breakdown of \model\/ to U(1)$_{\rm em}$
generates the charged $W$ boson mass matrix
$$
M_{W^\pm}^2 = \left( \begin{array}{cc} \displaystyle \frac{g_L^2}{4}\,
    ( 2 v_L^2 + |v|^2 + |w|^2) & -g_L g_R v^* w /2 \\ 
-g_L g_R v w^* /2 & \displaystyle
\frac{g_R^2}{4}\, (2 v_R^2 + |v|^2 + |w|^2)\end{array}
\right) \equiv \left( \begin{array}{cc} M_L^2 & M_{LR}^2\,e^{-i\lambda}
    \\ M_{LR}^2\,e^{i\lambda} & M_R^2\end{array}\right).
$$
The eigenvalues
\begin{eqnarray*}
M_1^2 & = & M_L^2 \cos^2\zeta + M_R^2 \sin^2 \zeta + M_{LR}^2 \sin
2\zeta,\\
M_2^2 & = & M_L^2 \sin^2\zeta + M_R^2 \cos^2 \zeta - M_{LR}^2 \sin
2\zeta,
\end{eqnarray*}
and eigenvectors
\addtolength{\arraycolsep}{3pt}
$$
\left( \begin{array}{c} W_1^+\\ W_2^+\end{array}\right) = \left[ 
\begin{array}{cc} \cos\zeta & -e^{i\lambda} \sin\zeta \\ 
e^{-i\lambda}\sin\zeta &
  \cos\zeta\end{array} \right] \left( \begin{array}{c} W_L^+\\ 
W_R^+\end{array}\right)
$$
\addtolength{\arraycolsep}{-3pt}
of this mass matrix correspond to the physical charged $W$ bosons in
the \model\/ model. The  $W_L$--$W_R$ mixing angle is defined as
$$
\tan 2\zeta = -\frac{2M_{LR}^2}{M_R^2-M_L^2}\, ,
$$
and the charged current reads (with $g\equiv g_L\equiv g_R$ and
without displaying unphysical scalars and charged Higgs contributions):
\begin{eqnarray*}
{\cal L}_{cc} & = & -\frac{g}{\sqrt{2}}\, \bar{U}_i \left[ \cos\zeta
  (V_L)_{ij} \gamma^\mu P_L - e^{-i\lambda} \sin \zeta (V_R)_{ij}
  \gamma^\mu P_R \right] D_j\, W^+_{1\mu}\\
& & {}-\frac{g}{\sqrt{2}}\, \bar{U}_i \left[ e^{i\lambda}\sin\zeta
  (V_L)_{ij} \gamma^\mu P_L + \cos \zeta (V_R)_{ij}
  \gamma^\mu P_R \right] D_j\, W^+_{2\mu}.
\end{eqnarray*}

\section{Quark Mixing}
\setcounter{equation}{0}

The Yukawa interaction part of the Lagrangian reads
\begin{equation}\label{eq:real}
-{\cal L}_Y = \Gamma_{ij} \bar q_{Li} \Phi q_{Rj} + \Delta_{ij} \bar
q_{Li} 
 \widetilde{\Phi} q_{Rj} + {\rm h.c.}
\end{equation}
with $\widetilde{\Phi} = \sigma_2 \Phi^* \sigma_2$.
As discussed in the introduction, the crucial feature of the SB--LR with
spontaneous CP violation is that P invariance 
coupled to spontaneous CP violation
restricts the coupling matrices $\Gamma$ and $\Delta$:
apart from one special case, see Ref.~\cite{EGNPB}, 
which we shall not consider here,
both matrices may  be taken real and
symmetric. We will work in this basis until further notice.
After spontaneous symmetry breaking one obtains the quark mass
matrices $M^{(u)}$ and $M^{(d)}$ of Eq.~(\ref{eq:MM}).
The diagonalisation of
$M^{(u)}$ normally requires a bi-unitary transformation, involving two
unitary matrices
acting separately on the L and R spinors; two more matrices are then
needed for  $M^{(d)}$.
In this special case however, the matrices $M^{(u)}$ and $M^{(d)}$
\emph{are symmetrical
in the chose basis} and can therefore be diagonalized by only two unitary
matrices $U,V$, so that
\begin{equation}\label{eq:UV}
M^{(u)} = U D^{(u)} U^T,\quad M^{(d)} = V D^{(d)} V^T,
\end{equation}
where  $D^{(u,d)}$ are diagonal mass matrices. Note that in the SB--LR the
signs of the quark masses are observable, so that the
entries in $D^{(u,d)}$ need in principle not be positive. 
This is a difference to the SM, where the sign of the
quark masses
in the Lagrangian,
$$
-m_i (\bar q_{Li} q_{Ri} + \bar q_{Ri} q_{Li}),
$$
can always be absorbed into the phase of  $q_R$, which is not observable,
since
right-handed quarks have no charged weak interactions. This is of course
no longer possible in the SB--LR.
For the moment, we thus need to keep track of  the possible signs of the
masses; they will later be absorbed into
the right quark mixing matrix so that the model has the 
 standard mass terms in the Lagrangian and the standard
quark propagator, $1/(p_\mu \gamma^\mu
-m)$, $m \geq 0$. 
As the CKM phases can only depend on mass ratios, we
thus have a $2^5=32\,$--fold multiplicity of solutions corresponding
to the different possible choices of the quark mass
signs. Fortunately, as we shall see later, phenomenological constraints
remove most of these solutions.

Diagonalizing the mass matrices introduces two quark mixing matrices,
one for the left, and one for the right sector; the crucial feature in
the present case of spontaneous CP violation is that in the special 
basis where $\Delta$ and $\Gamma$ are symmetrical, 
Eq.~(\ref{eq:UV}) implies that the left and right mixing matrices
 are complex
conjugate to each other:
$$
K \equiv K_L = U^\dagger V = K_R^*.
$$ 
The remainder of this section will be devoted to the calculation of
the phases of that matrix, but first we would like to make the
connection to a more ``standard'' quark basis in which the left mixing
matrix contains only one phase, $\delta$ 
(this choice is obviously not unique, and the numerous parametrisations
of the Kobayashi-Maskawa matrix attest to this; the procedure below is,
however,
general).
In order to do so, we first
rewrite $K$ as
$$
K = \zeta^u \tilde{K} \zeta^{d*},
$$
where $\zeta^{u,d}$ are diagonal matrices with entries $\zeta_i$ such
that $(\zeta_i)^2 = {\rm sign}\,(m_i)$. These phases are introduced to 
redefine the masses as positive. The advantage of introducing
$\tilde{K}$ is that for $\alpha=0$, i.e.\ for the case of no CP
violation,
and for a suitable choice of gauge, all entries are real.
Redefining now the phases of the quark fields by $\gamma^u_i$,
$\gamma^d_j$, $1\leq i,j\leq3$, one can bring $\tilde{K}$ into
a
standard form $V_L$ (i.e.\ with only one phase left):
$$
V_L = e^{-i\gamma^u} \tilde{K} e^{i\gamma^d}.
$$
The right mixing matrix then reads
$$
V_R = \eta^u e^{-i\gamma^u} \tilde{K}^* e^{i\gamma^d} \eta^d = 
\eta^u e^{-2 i\gamma^u}
V_L^* e^{2i\gamma^d} \eta^d.
$$
Here $\eta=\zeta^2$ are diagonal matrices with entries $\pm1$
and specify the signs of the quark masses.
Note that the phase
matrices introduce only five independent phases in $V_R$, as only
differences $\gamma^d_i-\gamma^u_j$ enter. $V_R$ can then be written as
\renewcommand{\arraystretch}{1.4}
\begin{equation}\label{eq:1}
V_R = \left(\begin{array}[c]{lll} 
\vlc{1}{1} e^{2 i \alpha_1} & \vlc{1}{2} e^{i
  (\alpha_1+\alpha_2+\epsilon_1)}& 
\vlc{1}{3} e^{i(\alpha_1+\alpha_3+\epsilon_1+\epsilon_2)}\\
\vlc{2}{1} e^{i(\alpha_1+\alpha_2-\epsilon_1)}& \vlc{2}{2} e^{2 i
  \alpha_2} & \vlc{2}{3} e^{i(\alpha_2+\alpha_3+\epsilon_2)}\\
\vlc{3}{1} e^{i(\alpha_1+\alpha_3-\epsilon_1-\epsilon_2)} & \vlc{3}{2}
e^{i(\alpha_2+\alpha_3-\epsilon_2)} & \vlc{3}{3}e^{2 i \alpha_3}
\end{array}\right)
\end{equation}
\renewcommand{\arraystretch}{1}
with the five independent phases $\alpha_i$, $\epsilon_i$, in which
we have also absorbed the signs of the quark masses; the 6th phase,
hidden in $(V_L^*)_{ij}$, is the usual unique surviving phase of the
SM model. 
Note that the phases of $V_L$ and $V_R$ depend on the parametrization
chosen for $V_L$.

We also would like to mention that in addition to different
parametrizations of $V_L$, also different conventions for the
phase in $W_L$--$W_R$ mixing are used in the literature. Most papers, 
notably \cite{langacker}
and \cite{EGNPB} (and ours), keep the phase explicitly in the 
Lagrangian, whereas \cite{JMF} prefers to shuffle it into $V_R$. 
This corresponds to a choice of gauge
$\lambda^{\mbox{\scriptsize\cite{JMF}}}=0$,
while REf.~\cite{EGNPB} uses $v = v^*$ instead, leading to 
$$e^{i\lambda^{[{\rm EG}]}} = - e^{-i\alpha}.$$
Denoting matrices in the different conventions by $V_L^{
\mbox{\scriptsize\cite{JMF}}}$
for
Ref.~\cite{JMF} and $V_L^{[{\rm EG}]}$ for Ref.~\cite{EGNPB}, we then
have
to identify
\begin{equation}\label{eq:2}
V_R^{\mbox{\scriptsize\cite{JMF}}} = e^{i\alpha} V_R^{[{\rm EG}]}.
\end{equation}
As later on we would like to use formulas
 given in Ref.~\cite{JMF}, we also have to
convert  the phases $\delta_1$, $\delta_2$ and
$\gamma$ of $V_R^{\mbox{\scriptsize\cite{JMF}}}$ into our
language. 
We find that one has to identify:
\begin{eqnarray*}
\gamma-\delta_2 & = & 2\alpha_1 + \alpha,\\
\gamma+\delta_2 & = & 2\alpha_2 + \alpha,\\
\gamma-\delta_1 & = & \alpha_1+\alpha_2+\epsilon_1 + \alpha,\\
\gamma+\delta_1 & = & \alpha_1+\alpha_2-\epsilon_1 + \alpha.
\end{eqnarray*}
Note that the system is degenerate and contains only three independent
 relations.

We can now parametrize the matrix $K$ as
\begin{equation}\label{eq:Kspecial}
\renewcommand{\arraystretch}{1.4}
K = \left(\begin{array}[c]{lll} 
\vl{1}{1} \,e^{-i \alpha_1} & \vl{1}{2} \,e^{-i/2\,
  (\alpha_1+\alpha_2+\epsilon_1)}& 
\vl{1}{3} \,e^{-i/2\,(\alpha_1+\alpha_3+\epsilon_1+\epsilon_2)}\\
\vl{2}{1} \,e^{-i/2\,(\alpha_1+\alpha_2-\epsilon_1)}& \vl{2}{2} \,e^{- i
  \alpha_2} & \vl{2}{3} \,e^{-i/2\,(\alpha_2+\alpha_3+\epsilon_2)}\\
\vl{3}{1} \,e^{-i/2\,(\alpha_1+\alpha_3-\epsilon_1-\epsilon_2)} &
\vl{3}{2}
\,e^{-i/2\,(\alpha_2+\alpha_3-\epsilon_2)} & \vl{3}{3}\,e^{- i \alpha_3}
\end{array}\right).
\end{equation}
\renewcommand{\arraystretch}{1}
It is clear that the phases will be functions of $r$ and $\alpha$, and
that CP violation is characterized by $y= r \sin\alpha$, with $r\sim
O(m_b/m_t)$. This fact led the authors of 
\cite{Chang,EGNPB} to calculate $\tilde{K}$ in a
linear expansion in $y$. Later on, in \cite{JMF}, the full solutions
for the phases for mixing between the first two generations were
calculated and it was found that the linear approximation works
perfectly well for these entries. It is, however, to be expected that
the
linear approximation breaks down for the third generation matrix
elements, where the natural ``smallness'' of the expansion parameter
$y$ can be upset by enhancement factors $m_t/m_b$, cf.\ also
Ref.~\cite{EGZPC}. In this paper 
we calculate the full phases beyond
the small phase approximation, which, as shown in \cite{JMF}, amounts
to solving the matrix equation
\begin{equation}\label{eq:matrix}
(1-r^2) \tilde{W}\, \tilde{D}^{(u)}\, \tilde{W} + 
(r^2 e^{i\alpha} - e^{-i\alpha}) 
\tilde{D}^{(u)} = 2 i r \sin\alpha \, \tilde{K}\, \tilde{D}^{(d)}\,
\tilde{K}^T,
\end{equation}
which is equivalent to a system of 12 (real) coupled equations.
The unknowns in these equations are six real parameters characterizing
the
unitary symmetrical matrix $\tilde{W}$ and the six phases of the mixing
matrix $\tilde{K}$. $\tilde{D}=\zeta\, D\,\zeta=\eta\, D$ 
are the diagonal mass matrices
including the signs of the quark masses. In order to solve
(\ref{eq:matrix}), it is
convenient to replace the independent variables $r$ and $\alpha$ by
new variables $\beta$ and $\beta'$, defined as \cite{JMF}
\begin{equation}\label{eq:def_beta}
\beta = \arctan \frac{2 r \sin\alpha}{1-r^2},\qquad e^{i\beta'} =
\frac{1-r^2 e^{-2i\alpha}}{|1-r^2 e^{-2i\alpha}|}.
\end{equation}
This transformation makes the dependence on one variable, $\beta'$,
trivial:
$$
\tilde{K} = e^{-i\beta'/2} K',
$$
where $K'$ is solution of
\begin{equation}\label{eq:basis}
\cos\beta\, W'\, \tilde{D}^{(u)}\, W' - \tilde{D}^{(u)} = 
i\sin\beta\, K'\, \tilde{D}^{(d)}\,K'^T,
\end{equation}
and $W'$ is still unitary and symmetric.
For the ``natural'' choice $r\sim O(m_b/m_t)$, to be motivated in the
next section, $\beta'$ is
negligibly small, and we shall neglect it in our final 
results.\footnote{$\beta'$ is, however, in general not negligible in the 
large-mixing region.}
The phases are thus to an excellent approximation functions of only one
variable, $\beta$.
As shown in Ref.~\cite{JMF}, the above equation has solutions only for a
restricted interval in $\beta$,
$$
\tan\,\frac{\beta}{2} \leq \frac{m_b}{m_t}.
$$
We have solved (\ref{eq:basis}) by a polynomial expansion as
suggested in \cite{JMF}. The input parameters are the quark masses
renormalized at
one common scale which we choose to be $\bar m_t$; we use the
following values, in the $\overline{\rm MS}$ scheme:
\begin{equation}
\renewcommand{\arraystretch}{1.4}
\begin{array}[b]{lllll}
\bar m_t(\bar m_t) & = & 170\,{\rm GeV},\\
\bar m_b(\bar m_t) & = & 2.78\,{\rm GeV} & \Leftrightarrow & 
\bar m_b(\bar m_b) = 4.25\,{\rm GeV},\\
\bar m_c(\bar m_t) & = & 0.63\,{\rm GeV} & \Leftrightarrow & 
\bar m_c(\bar m_c) = 1.33\,{\rm GeV},\\
\bar m_s(\bar m_t) & = & 0.060\,{\rm GeV} & \Leftrightarrow & 
\bar m_s(2\,{\rm GeV}) = 110\,{\rm MeV},\\
\bar m_d(\bar m_t) & = & 0.0030\,{\rm GeV} & \Leftrightarrow & m_s/m_d =
20.1,\\
\bar m_u(\bar m_t) & = & 0.0017\,{\rm GeV} & \Leftrightarrow & m_u/m_d =
0.56.
\end{array}
\renewcommand{\arraystretch}{1}
\end{equation}
The value of $\bar m_s$ is a compromise between recent lattice
calculations as summarized in \cite{sinead} and QCD sum rule
calculations \cite{msSRs}.

One more input are the angles of the CKM matrix. The particle data
group \cite{PDG} quotes for the entries determined from tree-level processes,
which receive only tiny corrections from LRS contributions:
\begin{eqnarray}
|V_{ud}| & = & 0.9740\pm 0.0010,\qquad  
|V_{us}| = 0.2196 \pm 0.0023,\nonumber\\
|V_{cb}| & = & 0.0395\pm 0.0017,\qquad |V_{ub}/V_{cb}| = 0.08\pm 0.02.
\label{eq:exp}
\end{eqnarray}
For an exactly unitary matrix where all entries lie within the above
specified error range, we fix
\begin{equation}\label{eq:fix}
|V_{us}| = 0.2219,\quad |V_{ub}| = 0.004, \quad |V_{cb}| = 0.04.
\end{equation}
In addition, we have to specify the value of the phase $\delta$ for
the case of no CP violation, i.e.\ $\delta(\beta=0)=0$ or
$\pi$. We will label the corresponding set of solutions of
(\ref{eq:basis}) as class I and class II solutions, respectively.
This induces another two-fold multiplicity in addition to the
32-fold one from the different quark mass signs, so that we finally
have to deal with 64 different solutions of (\ref{eq:basis}). In
Tab.~\ref{tab:1} we give the explicit identifications of the 
solutions with the mass signatures.
\begin{table}
\begin{center}
\renewcommand{\arraystretch}{1.2}
\begin{tabular}{r|cccccc}
\hline\hline
no. & $m_t$ & $m_b$ & $m_c$ & $m_s$ & $m_d$ & $m_u$\\ \hline
1 & + & + & + & + & + & +\\
2 & + & + & + & + & -- & +\\
3 & + & + & + & -- & + & +\\
4 & + & + & + & -- & -- & +\\
5& + & + & -- & + & + & +\\
6& + & + & -- & + & -- & +\\
7& + & + & -- & -- & + & +\\
8& + & + & -- & -- & -- & +\\
9& + & -- & + & + & + & +\\
10& + & -- & + & + & -- & +\\
11& + & -- & + & -- & + & +\\
12& + & -- & + & -- & -- & +\\
13& + & -- & -- & + & + & +\\
14& + & -- & -- & + & -- & +\\
15& + & -- & -- & -- & + & +\\
16& + & -- & -- & -- & -- & +\\
17& -- & + & + & + & + & +\\
18& -- & + & + & + & -- & +\\
19& -- & + & + & -- & + & +\\
20& -- & + & + & -- & -- & +\\
21& -- & + & -- & + & + & +\\
22& -- & + & -- & + & -- & +\\
23& -- & + & -- & -- & + & +\\
24& -- & + & -- & -- & -- & +\\
25& -- & -- & + & + & + & +\\
26& -- & -- & + & + & -- & +\\
27& -- & -- & + & -- & + & +\\
28& -- & -- & + & -- & -- & +\\
29& -- & -- & -- & + & + & +\\
30& -- & -- & -- & + & -- & +\\
31& -- & -- & -- & -- & + & +\\
32& -- & -- & -- & -- & -- & +\\\hline\hline
\end{tabular}
\end{center}
\renewcommand{\arraystretch}{1}
\caption[]{Identification of solutions of
  (\protect{\ref{eq:matrix}}) with quark mass signatures. We choose
  $m_u$ to be always positive.}\label{tab:1}
\end{table} 
\begin{figure}[p]
\vspace*{-0.5cm}
$$
\epsffile{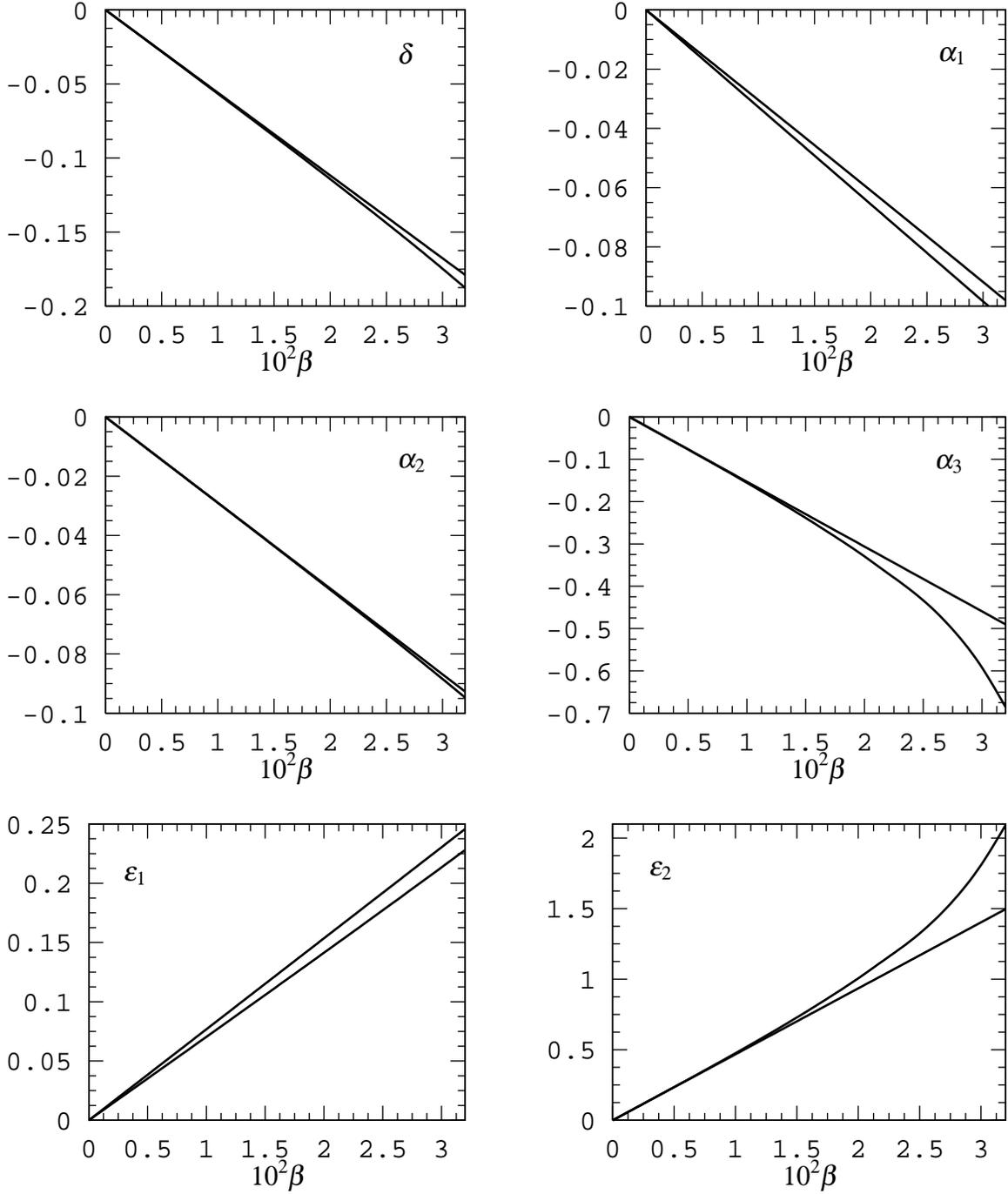}
$$
\caption[]{Independent phases of the CKM matrices acc.\ to
  (\protect{\ref{eq:1}}) as functions of $\beta$ in the Maiani convention, 
for $\delta(\beta = 0)=0$, i.e.\ class I, and $\beta'=0$,
for positive quark masses. The straight lines are the phases
calculated in the small phase approximation, the curves are the full
results.}\label{fig:phases}
\end{figure}

Before presenting results, we would also like to stress that $K'$ and
$\tilde{K}$ are independent of the phase-convention for $V_L$ -- they
only depend on the modulus $|V_L|$. The phase-convention for $V_L$
enters just
in the extraction of the six phases $\delta$, $\alpha_i$
and $\epsilon_i$ from Eq.~(\ref{eq:Kspecial}). In the following, we
will always use the Maiani convention
$$\addtolength{\arraycolsep}{1pt}
V_L = \left( \begin{array}{ccc}
c_{12}c_{13} & s_{12} c_{13} & s_{13} e^{-i\delta}\\
-s_{12}c_{23}-c_{12} s_{23} s_{13} e^{i\delta} &
c_{12}c_{23}-s_{12}s_{23}s_{13} e^{i\delta} & s_{23}c_{13}\\
s_{12}s_{23}-c_{12}c_{23}s_{13}e^{i\delta} &
-c_{12}s_{23}-s_{12}c_{23}s_{13} e^{i\delta} & c_{23} c_{13}
\end{array}\right),
\addtolength{\arraycolsep}{1pt}
$$
with, from the experimental results, Eq.~(\ref{eq:exp}),
$$
\theta_{12} = 0.2218\pm 0.0031, \quad \theta_{23} = 0.0395\pm 0.0017,
\quad \theta_{13} = 0.0032 \pm 0.0008.
$$
Our preferred values for CKM matrix elements (\ref{eq:fix}) correspond
to
\begin{equation}\label{eq:CKMfix}
\theta_{12} = 0.2238, \quad \theta_{23} = 0.0400, \quad \theta_{13} =
0.004,
\end{equation}
so that\addtolength{\arraycolsep}{2pt}
$$
V_L (\delta = 0) = \left(\begin{array}{rrr}
0.9751 & 0.2219 & \phantom{-}0.0040\\ -0.2219 & 0.9743 & 0.0400\\
0.0050 & -0.0399 & 0.9992\\
\end{array}\right)
$$
and
$$
V_L (\delta = \pi) = \left(\begin{array}{rrr}
0.9751 & 0.2219 & -0.0040\\ -0.2216 & 0.9743 & \phantom{-}0.0400\\
0.0128 & -0.0381 & 0.9992\\
\end{array}\right).
$$\addtolength{\arraycolsep}{-2pt}
Note that $|V_{td}|$ is quite sensitive to the value of $\delta$.  A
full analysis of the impact of 
  CKM angle and quark mass uncertainties on our results is
  beyond the scope of this paper.

In Fig.~\ref{fig:phases} we plot the independent phases as
functions of $\beta$ in the allowed range $0\leq \beta \leq 2 m_b/m_t =
0.0327$, both the full results and the small phase approximation, for
$\delta(\beta = 0) = 0$, i.e.\ class I, and positive quark masses.
{}From the figure it is evident that the two phases characteristic for
the 3rd generation, $\alpha_3$ and $\epsilon_2$, deviate significantly
from the small phase approximation for large $\beta$, 
whereas for the other phases the
agreement between the full and the approximate result is very
good. This feature is also observed for the other 63 solutions. It is
also
evident that the phases characterizing mixing between the first two
generations are rather smallish, whereas $\alpha_3$ and $\epsilon_2$
can become large. This reflects the impact of enhancement factors
$\sim m_t/m_b$ that overcome the smallness of $\beta$ and invalidate
the small phase approximation. The numerical results for all the 64
\begin{figure}[t]
$$
\epsffile{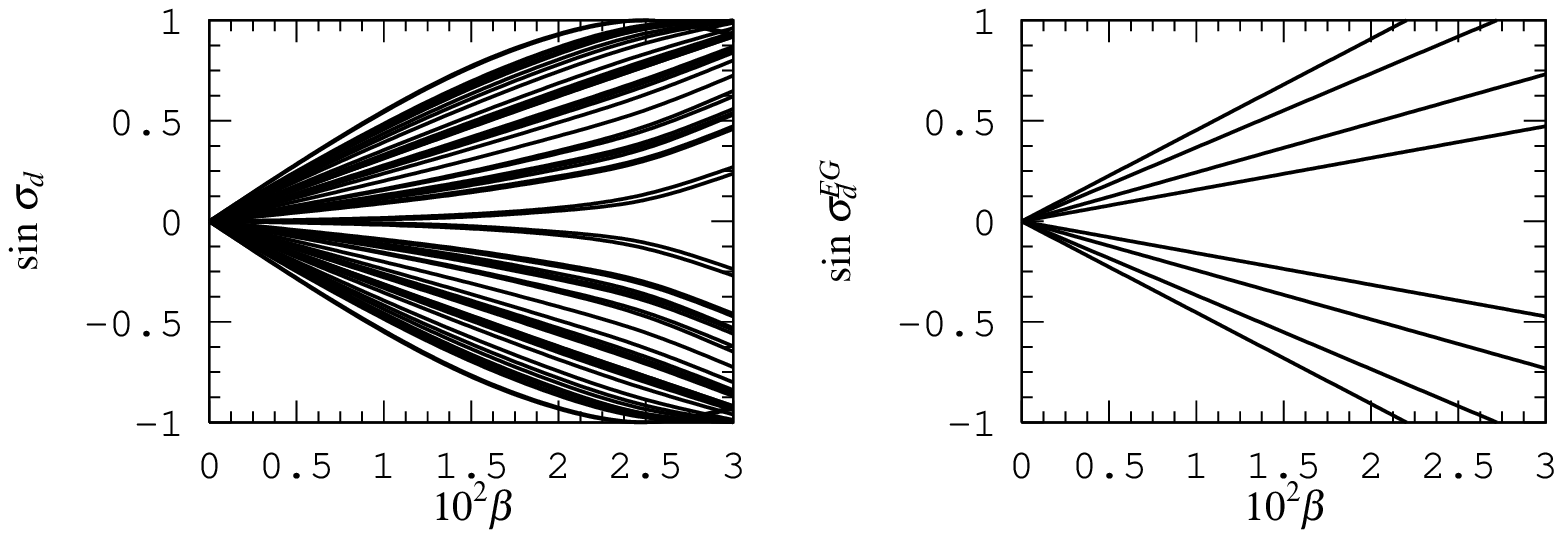}
$$
\vspace*{-30pt}
\caption[]{$\sin \sigma_d$, defined in (\protect{\ref{eq:def_sigma}}), 
  as function of
  $\beta$ calculated from the full solution, left, and in the
  approximation derived in Ref.~\protect{\cite{EGZPC}}, right. Note that
  with the approximate formula, the value of the sinus can become
  larger than 1 for certain choices
  of the mass signs.}\label{fig:sigdEG}
$$
\epsffile{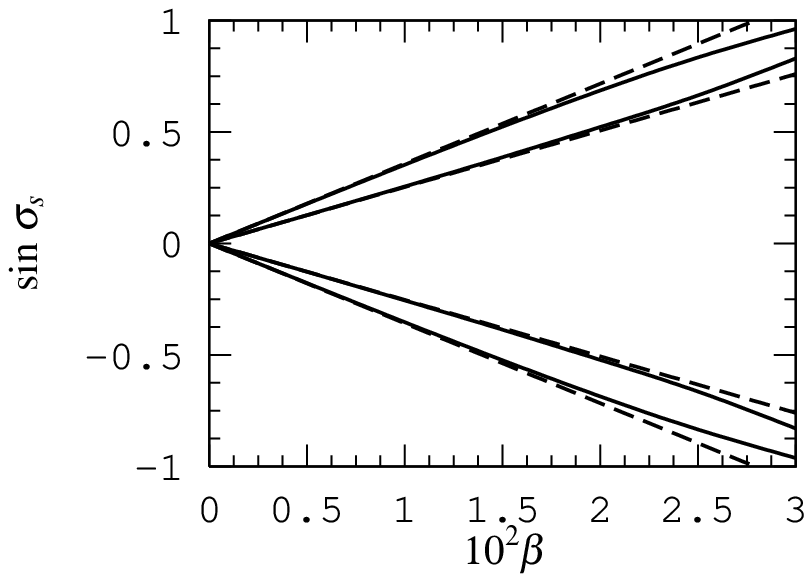}
$$
\vspace*{-30pt}
\caption[]{Same as previous figure, but for $\sin \sigma_s$. Dashed
  lines are the small phase approximation.}\label{fig:sigsEG}
\end{figure}
solutions are available from the authors as {\sc Mathematica} file.

Finally, we would like to comment on the approximate formulas for
the sinus of 
certain parametrization-independent 
combinations of CKM entries relevant for B mixing, 
$\sigma_d$ and $\sigma_s$, to be
defined in Sec.~5, which were originally derived in \cite{EGZPC} and
 used in \cite{bloedsinn} and follow-up papers. These approximate
formulas rely on the small phase approximation and retain only leading
terms in the ratios of quark masses. In Fig.~\ref{fig:sigdEG} we plot
$\sin \sigma_d$ as function of $\beta$ calculated from the full
solutions and from the approximate formula given in
\cite{EGZPC}. Apparently, the structure of the full solution is richer
than can be reproduced by a simple approximation formula. Also, 
for a number of quark mass signatures and large $\beta$, the formula for
$\sin \sigma_d$ predicts values larger than 1 and thus cannot be used in
that region.  On the other hand, for $\sin \sigma_s$, 
shown in Fig.~\ref{fig:sigsEG}, the approximation
works quite well and only fails for large $\beta$. 

\section{The SB--LR Parameter Space and Strategies for Constraints}
\setcounter{equation}{0}

The new parameters of the SB--LR are the following (numerical 
values will be discussed in later sections):
\begin{itemize}
\item $M_2\sim O(1\,{\rm TeV})$, the mass of the predominantly right-handed 
weak gauge boson;
\item $\zeta$, the mixing angle between $W_R$ and $W_L$, $\zeta\geq 0$;
\item $g_R$, the coupling of $W_R$. As discussed in the introduction,
although low-energy parity restoration would require $g_R = g_L$, this 
is not necessarily requested if parity, but not \model\ is broken
at a higher scale, and $g_R \neq g_L$ may be preferable to avoid 
domain-wall formation. For definiteness we will however set here
$g_R = g_L$, as the results on boson masses can be easily adapted;
\item $0\leq r\leq 1$ and $0\leq \alpha\leq \pi$, parametrizing 
the spontaneous breakdown of 
CP symmetry, i.e.\ the VEV of the bidoublet $\Phi$;
\item extra Higgs masses which,
in principle, are quite arbitrary. However, since they
are associated with neutral flavour changing currents, the prejudice
is usually to have heavy extra Higgses with masses $M_H\sim O(10\,{\rm TeV})$.
In this case, they cannot mix significantly with the lighter ones
and must be nearly degenerate (their splitting being at most
of the order of the weak scale), see also \cite{EGNPB}. We will in the
present study neglect the mass differences, and allow only one single 
mass parameter for the heavy Higgs bosons, $M_H$, and also assume $M_H>M_2$.
A full study, allowing extra Higgses lighter than the
 right-handed bosons could some day be needed; it would however be sensitive
 to the fine details of the scalar potential.
\end{itemize}
Actually not all of these parameters are independent, but they observe the 
following relations and constraints:
\begin{eqnarray}
\zeta & = & \left( \frac{m_W}{M_2}\right)^2 \frac{2r}{1+r^2},\label{eq:zeta}\\
\frac{m_b}{m_t} & > & \left| \frac{r
    \sin\alpha}{1-r^2}\right|,\label{eq:ra}\\
\frac{M_H}{M_2} & < & 13.\label{eq:panzerknacker}
\end{eqnarray}
The second of these relations originates from quark mixing, see
Sec.~3, and the third one comes from requiring convergence of the
perturbation expansion \cite{higgsbound}.

Let us now discuss for what type of processes we expect measurable
effects due to the SB--LR. First of all, SB--LR implies strong
constraints on the CKM phase $\delta$ even in the decoupling limit of the
model, i.e.\ $M_2,M_H\to\infty$. An inspection of all sets of quark
mixing phases shows that $|\delta^{\rm SB-LR}|<0.25$ for class I
solutions and $|\delta^{\rm SB-LR}-\pi|<0.25$ for class II
solutions. One can make use of this fact to exclude the SM model limit
of the SB--LR model experimentally by measuring $\delta$ from 
$\Delta m_{B_d}$ or $\epsilon$. The extraction of $\delta$ from
these measurements, however, 
to date involves
considerable theoretical uncertainties. Nevertheless, a recent global
fit of SM CKM parameters \cite{fit}, which includes conservative
theory error estimates, finds $\delta^{\rm SM} = 1.0\pm
0.2$, so that we conclude that {\em the decoupling limit $M_2,M_H\to\infty$ of
  the SB--LR is excluded by 3.5$\,\sigma$}. 

Let us next discuss the
case of finite $M_2$ and $M_H$.
As for $M_2$, it enters
either directly via a $W_2$ propagator or indirectly in $\zeta$ via
$W_L$--$W_R$ mixing. Obviously, with an expected modification of the
amplitude of size $(m_W/M_2)^2\sim
O(10^{-3})$ it is in practive impossible to observe either of these
effects in tree-level decays. In loop-induced processes, however, the
situation changes, and we expect measurable or even sizable effects for
the following cases:
\begin{itemize}
\item the suppression factor $(m_W/M_2)^2$ is partially compensated by
  large matching coefficient functions (tree-level
  Wilson-coefficients) of the effective
  Hamiltonian, radiative corrections or hadronic matrix elements
  (e.g.\ chiral enhancement in K mixing\footnote{The fact that there
    is no such chiral enhancement for B mixing led some authors to
    conclude that the SB--LR would not have much impact on these
    processes, cf.\ e.g.\ \cite{GL}; 
    indeed, it is the Higgs contribution that is dominant
    in B mixing.});
\item $W_L$--$W_R$ mixing is enhanced by large quark mass terms from
  spin-flips, e.g.\ $\zeta\to \zeta\, m_t/m_b$ in $b\to s\gamma$
  \cite{Misiak}; this affects all top-dominated penguin-diagrams and
   is thus expected to be important for processes with
  direct CP violation;
\item the SM amplitude is forbidden or heavily suppressed (electric
  dipole moment of the neutron).
\end{itemize}

As for the Higgses, their contribution to SM tree-level decays is also
heavily suppressed by factors $(m_W/M_H)^2\sim 10^{-4}$ or smaller
from the propagator. On the other hand, the neutral FC Higgs contributes to
$\Delta F=2$ processes at tree-level and the charged Higgs
contributions get enhanced by
$m_t/m_b$ in $b$ penguins; roughly speaking, their contribution is
similar in size to that of $W_R$ for K mixing and becomes dominant in
B mixing.

Based on these relations, we distinguish two regions in 
parameter space with qualitatively different 
phenomenological consequences. These are
\begin{itemize}
\item the ``natural'' region with $r\sim O(m_b/m_t)\sim 0.02$, which implies
$\zeta \sim O(10^{-4})$;
\item the ``large mixing'' region with $\zeta\sim 10^{-3}$ so that
$r\sim (0.1$--$1)$.
\end{itemize}
The small $r$ region is called ``natural'', because it has been argued
to explain the observed 
smallness of the CKM mixing angles \cite{natural}. 
With the expected size of $M_2\sim O(1\,{\rm TeV})$, one
then has a rather small mixing angle $\zeta\sim 10^{-4}$, which severely 
restricts the possible impact of SB--LR contributions on penguin-induced
 processes and on CP asymmetries from direct CP violation. On the
 other hand, condition
(\ref{eq:ra}) is fulfilled for any $\alpha$, which can thus vary freely 
within $0$ and $\pi$. In the natural region, $\zeta$ effectively 
decouples from mixing-induced CP violating processes and we can choose $r$
 and $M_2$ as independent variables.

In the large mixing region, on the other hand, we {\em require} $\zeta$ to be
close to its maximum experimentally allowed value $\zeta \approx 0.003$ 
\cite{langacker,JMF}. Now $r$ can become large, and consequently, via
(\ref{eq:ra}), $\alpha$ is restricted to values close to $0$ or $\pi$. This is 
the region where one might expect sizable SB--LR effects to show up in
penguin-induced  
processes. In this case it is more appropriate to choose $\zeta$ and $M_2$ 
as independent variables and determine $r$ from Eq.~(\ref{eq:zeta}). 

In the present paper we restrict ourselves to mixing-induced CP violating 
effects and thus work consistently in the natural region, fixing\footnote{
As long as $r\leq m_b/m_t$, its exact value does not matter.} 
$r=m_b/m_t$. The remaining independent 
parameters 
are then $\beta$, $M_2$ and $M_H$. In addition, we have a 64-fold 
multiplicity of CKM phases from the different possible choices for quark
mass signs and the value of the phase $\delta$ in the limit of no CP 
violation, $\delta=0$ or $\pi$. The observables we analyse in this
paper are $\Delta m_K$, $\Delta m_{B_{d,s}}$, $a_{CP}(B\to J/\psi
K_S)$, $\epsilon$ and $\epsilon'/\epsilon$. 
Another possible observable from the $B_s$ system is $\Delta\Gamma_s$,
which was analyzed in \cite{xx}.
Another potentially powerful constraint can in principle be obtained 
 from the neutron
electric dipole moment (EDM). The LR model contributions to the EDM are
discussed in Refs.~\cite{JMF} and \cite{JMF2}, taking into account
not only the sum of the quark EDM's as done in previous calculations, 
but also specific 
hadronic terms which involve $W_L$--$W_R$ exchange between the 
quark lines of a neutron. It is indeed an interesting 
feature of the SB--LR model that it allows CP violation 
in this sector already within a 1-generation context.
Crucial to such contributions is the presence of LR mixing,
since the EDM is basically a LR transition and hence suppressed
in the SM.
 In Ref.~\cite{JMF2}, the following bound was obtained:
$$
\left|\vphantom{\frac{2r}{1+r^2}}\zeta  \sin(\gamma - \delta_2)\right|= 
\frac{2r}{1+r^2}\left(\frac{m_W}{M_2}\right)^2
\left|\vphantom{\frac{2r}{1+r^2}} \sin\left\{2\alpha_1(\alpha) +
\alpha\right\}\right| \leq 3 \cdot 10^{-6}.
$$
Yet, this bound is not without criticism: it is obtained as a sum of
large terms of opposite sign and thus comes with considerable
uncertainty. In addition, it is to be supposed that there are also
large contributions from gluonic matrix elements, i.e.\ strong CP
violation, which might upset the bound. We thus refrain from taking it
into account in our analysis.

Most of the obervables we analyze in this paper are related to
the matrix element
\begin{equation}
\langle M^0 | {\cal H}^{|\Delta| F=2}_{\rm eff}| \bar{M}^0\rangle = 
2 m_M \left(M_{12}^{\rm SM} + M_{12}^{\rm LR} + M_{12}^{\rm LD}\right),
\end{equation}
$F=S,B$ and $M^0 = K^0$, $B_d^0$ and $B_s^0$. $M_{12}^{\rm SM}$ stands
for the SM contribution, $M_{12}^{\rm LR}$ for the SB--LR contribution
and $M_{12}^{\rm LD}$ for $(\Delta F=1)^2$ 
contributions, which are negligible in
B mixing, but expected to be sizable in
K mixing \cite{df2}. 
We also introduce the mixing angles $\phi_M^{B_q}$ and $\phi_M^K$,
\begin{equation}\label{eq:phi}
 \phi^{B_q}_M = {\rm arg}\,
M_{12}^{B_q},\qquad  \phi^{K}_M = {\rm arg}\, M_{12}^K.
\end{equation}
In both the B and the K system the mixing between flavour eigenstates
can be described in terms of the these mixing angles to an excellent accuracy; 
the only quantities to be considered in this paper, for which that 
approximation
is not sufficient, are $\epsilon$ and $\epsilon'/\epsilon$.
Note also that the mixing angles are
convention-dependent quantities.

As for CP conserving quantities, constraints can be derived from  $\Delta
m$, the mass difference between  mass eigenstates.
For both K's and B's, one has
$$
\Delta m = 2 \left| M_{12} \right|.
$$
The experimental mass differences in the K and B
system provide in principle a powerful constraint 
on $|M_{12}^{\rm LR}|$; for the K system,
however, the size of long-distance contributions to $M_{12}^{\rm LD}$
is not very well
known, so in this case one usually makes the reasonable 
  assumption\footnote{This assumption appears reasonable unless there
  are  large cancellations between the different
  contributions to $M_{12}$.} that the LR contribution should at most 
saturate $\Delta m_K$. We thus constrain the LR parameters by
requiring 
\begin{equation}\label{eq:wuerg}
2 |M_{12}^{\rm K,LR}| < \Delta m_K^{\rm  exp}.
\end{equation}
For $\Delta m_{B_d}$, as
LD contributions are negligible, theory is in a better shape and we
can require
\begin{equation}
2 |M_{12}^{\rm B}|  = \Delta m_{B_d}^{\rm
  exp}.
\end{equation}

We also investigate the CP violating observables $a_{\rm CP}(B\to J/\psi
K_S)$, to be defined in Sec.~5.2, which depends on both the K and B
mixing angles, and $\epsilon$ and $\epsilon'/\epsilon$ for the K
system, which will be investigated in Sec.~6. These three observables 
vanish for $\beta\to 0$, in contrast to $\Delta m_{K,B}$. {}From the
analysis of $\epsilon$ in particular, we shall conclude that the
SB--LR is excluded in the decoupling limit $M_2,M_H\to\infty$.

\section{B$^0$--$\bar{\mbox{\rm\bf B}}^0$ Mixing in the SB--LR}
\setcounter{equation}{0}

In this section we follow largely the notations and conventions of
Ref.~\cite{babar}.

\subsection{Constraints from $\Delta m_{B_d}$ and Predictions for 
$\Delta m_{B_s}$}

In the SM, $M_{12}$ is dominated  by box-diagrams with $W_L$ and top
exchange and given by
\begin{eqnarray}
M_{12}^{\rm SM} & = & \frac{1}{32\pi^2m_B}\, G_F^2 m_W^2 
(\lambda_t^{LL})^2 S(x_t)\eta^B_2(\mu)
\langle B^0 | [(\bar d b)_{V-A} (\bar d b)_{V-A}](\mu) | \bar
B^0 \rangle\nonumber\\
& \equiv & \frac{1}{32\pi^2m_B}\, G_F^2 m_W^2 
(\lambda_t^{LL})^2 S(x_t)\hat\eta^B_2(\mu) [\alpha_s^{(5)}(\mu)]^{-6/23}\,
\langle B^0 | [(\bar d b)_{V-A} (\bar d b)_{V-A}](\mu) | \bar
B^0 \rangle.
\end{eqnarray}
The hadronic matrix element is parametrized as usual as
\begin{equation}
\langle B^0 | [(\bar d b)_{V-A} (\bar d b)_{V-A}](\mu) | \bar
B^0 \rangle = - 2\left( 1 + \frac{1}{N_c}\right) \, B_B(\mu) f_B^2 m_B^2
e^{-i\phi^B_{CP}},\label{eq:Bfac}
\end{equation}
where the bag-factor $B_B(\mu)$ describes the deviation from vacuum saturation
 ($B_B^{\rm vac} = 1$) and $\phi^B_{CP}$ is an arbitrary phase describing the 
transformation behaviour of B flavour eigenstates under CP transformations:
$$
{\rm (CP)} |B^0\rangle =  e^{i\phi^B_{CP}} |\bar B^0\rangle, \quad 
{\rm (CP)} |\bar B^0\rangle\ =\ e^{-i\phi^B_{CP}} |B^0\rangle.
$$
It goes without saying that physical observables must be independent of that 
phase.
Often, the tacit choice $\phi^B_{CP} = \pi$ is made. $f_B$ is the 
leptonic decay constant of the B meson, defined as  
$$
\langle 0 | (\bar d b)_A | \bar B^0\rangle = i f_B p_\mu,
$$
and the Inami--Lim function $S$ is defined as
$$
S(x_t) = \frac{4x_t-11x_t^2+x_t^3}{4(x_t-1)^2} - \frac{3}{2}\,
  \frac{x_t^3}{(1-x_t)^3}\, \ln x_t
$$
with $x_t = \bar m_t(\bar m_t)^2/m_W^2$. For the CKM factors, we use the 
notation
$$
\lambda_t^{AB} = V^{A*}_{tq} V^{B}_{tb}
$$
with $A,B = L,R$. Numerical values of input parameters are given in 
Table~\ref{tab:x}.
\begin{table}
\renewcommand{\arraystretch}{1.5}
\addtolength{\arraycolsep}{3pt}
$$
\begin{array}{ccccccccc}
\hline\hline
\bar m_t(\bar m_t) & \eta_2^{B,LO}(m_b) & \hat{\eta}_2^{B,LO} & 
\hat{\eta}_2^{B,NLO} & B_B(m_b) & \hat{B}_B^{NLO} & f_B\sqrt{\hat{B}_B^{NLO}}
\\\hline
(170\pm 5)\,{\rm GeV} & 0.86 & 0.580 & 0.551 & 0.9\pm 0.1 & 1.4\pm 0.1 & 
(207\pm 42) \,{\rm MeV} \\\hline\hline
\end{array}
$$
\caption[]{Theoretical input parameters for B mixing. Numbers are taken from 
\protect{\cite{buhmann,laurent}}.}\label{tab:x}
$$
\begin{array}{cc}
\hline\hline
f_{B_s}/f_B & B_{B_s}/B_B\\\hline
1.14 \pm 0.08 & 1.00\pm 0.03\\\hline\hline
\end{array}
$$
\renewcommand{\arraystretch}{1.5}
\addtolength{\arraycolsep}{3pt}
\caption[]{Matrix elements for $B_s$, from \protect{\cite{laurent}}.
}\label{tab:su3}
\end{table}
In the SB--LR, there are several additional contributions, notably the
tree-level neutral Higgs exchange
and box-diagrams with $W_R$ and unphysical
scalar exchanges, all of which are dominated
by the top quark. 
Taking into account only the leading
contributions in $M_2$ and $M_H$, one finds:\footnote{Box-diagrams
  with charged Higgses are
  suppressed relative to the L--R box-diagrams by roughly a factor
 $(M_2/M_H)^2$ and thus can be neglected as long as
  $M_2\ll M_H$. Note also that $M_{12}^{\rm LR}$ as given in
  (\ref{eq:x}) is gauge-dependent; subsequent formulas are given in the
  t'Hooft-Feldman gauge. The diagrams restoring gauge-invariance have
  been calculated in \cite{soni,gauge} and were found to be small in
  that gauge.}
\begin{eqnarray}
M_{12} & = & M_{12}^{\rm SM} + M_{12}^{W_1W_2} + M_{12}^{S_1W_2} +
M_{12}^H\label{eq:x}\\
& \equiv & M_{12}^{\rm SM} + M_{12}^{\rm LR}\nonumber
\end{eqnarray}
with \cite{EGZPC}
\begin{eqnarray*}
M_{12}^H & = & -\frac{\sqrt{2}G_F}{m_BM_H^2}\, \bar{m}_t(\bar{m}_t)^2
\eta^H(\mu) \lambda_t^{LR} \lambda_t^{RL} \langle B^0 | O_S(\mu) |
\bar B^0\rangle,\\
M_{12}^{W_1W_2 + S_1W_2} & \simeq & -\frac{G_F^2}{8\pi^2}
\frac{m_W^4}{M_2^2} \, B_B^S(\mu) f_B^2 m_B \left[ 
     \frac{m_B^2}{\bar{m}_b(\bar{m}_b)^2}
   + \frac{1}{6} \right] \lambda_t^{LR} \lambda_t^{RL} \\
& &\times \left\{ 4
   \eta_1^{LR}(\mu) F_1(x_t,\beta) - \eta_2^{LR}(\mu)
   F_2(x_t,\beta) \right\} e^{-i\phi^B_{CP}}
\end{eqnarray*}
with $\beta = m_W^2/M_2^2$.

Let us first discuss the Higgs contribution.
To leading logarithmic accuracy, the operator $O_S = (\bar d P_L b)
(\bar d P_R b)$ renormalizes multiplicatively with an anomalous
dimension that just compensates that of the factor $m_t^2$ 
such that $m_t^2 O_S$ is RG-invariant. The LO short-distance
correction
$\eta^H$ can thus be written as
$$
\eta^H(\mu) = \left
  [ \frac{\alpha_s(\mu)}{\alpha_s(m_t)}\right]^{24/23}, \quad
\eta^H(m_b) = 2.0.
$$
On the other hand, the matrix element of $O_S$ can be written
as\footnote{
Note that our expressions for $M_{12}^{\rm LR}$ and the
matrix element of $O_S$ are by a factor
of 2 smaller than those quoted in \cite{EGZPC}, which is due to the
fact that the authors of that paper use a definition of $f_B$ which
makes
it by a factor of $\sqrt{2}$ smaller than ours. This is also the 
source of a factor of 2 discrepancy in 
the expression for $|M_{12}^{LR}/M_{12}^{SM}|$ in 
Ref.~\cite{bloedsinn}.}
 \begin{equation}\label{eq:OS}
\langle B^0 | O_S(\mu) | \bar B^0\rangle = -\frac{1}{2}\, f_B^2 m_B^2
B_B^S(\mu) \left[ \frac{m_B^2}{\bar m_b(\bar m_b)^2} +
  \frac{1}{2N_c}\right] e^{-i \phi^B_{CP}},
\end{equation}
where the bag--factor $B_B^S(\mu) \sim O(1),\ \mu\sim O(m_b)$, 
 contains the full scale-dependence
of the matrix-element. To our knowledge, $B_B^S(m_b)$ has never been
estimated by any non-perturbative method. In the appendix we calculate
the ratio $B_B^S/B_B$ both to leading order in a $1/N_c$ expansion and
from QCD sum rules. The results agree with
\begin{equation}\label{eq:Bes}
\frac{B_B^S(m_b)}{B_B(m_b)} = 1.2 \pm 0.2.
\end{equation}
 Motivated by the results in Tab.~\ref{tab:su3}, which indicate a
 small 
SU(3)-breaking for the bag-factor in the SM, 
we will use (\ref{eq:Bes}) also for the
 matrix elements over $B_s^0$.

As for $M_{12}^{W_1W_2 + S_1 W_2}$, its operator structure is more
complicated than that of the Higgs contribution. The LO short-distance
corrections $\eta_{1,2}^{LR}$ have been calculated in an approach 
suggested by Novikov, Shifman,
Vainshtein and Zakharov \cite{V}, which in Ref.~\cite{paschos} was shown to
be equivalent to the by now standard effective theory approach.
 {}From the results in
\cite{EGNPB,EGZPC}, one finds
\begin{equation}\label{eq:etas}
\eta_1^{LR}(m_b) \approx 1.8,\qquad \eta_2^{LR}(m_b) \approx 1.7.
\end{equation}
 $F_1$ and $F_2$ are in general
complicated functions of $x_t$, $x_b$ and $\beta =
(m_W/M_2)^2$. However, in
the limit $x_b\to 0$ and for $M_2\geq1.4\,$TeV, they are to within 5\%
accuracy approximated by
\begin{eqnarray*}
F_1 & \simeq & \frac{x_t}{1-x_t} + \frac{x_t\ln x_t}{(1-x_t)^2} - x_t
\beta \ln \beta,\\
F_2 & \simeq & \frac{x_t^2}{1-x_t} + \frac{2-x_t}{(1-x_t)^2}\, x_t^2
\ln x_t - x_t \ln \beta.
\end{eqnarray*}
Numerically, $|4F_1|\ll |F_2|$. We thus approximate  
$$
 \eta_2^{LR} F_2 - 4 \eta_1^{LR} F_1 \approx  \eta_2^{LR} ( F_2 - 4
F_1).
$$

We are now in a position to calculate $M_{12}$ in the SB--LR and to
investigate its impact on phenomenology. Following \cite{EGZPC}, we
write (\ref{eq:x}) as
\begin{eqnarray}
M_{12} & = & M_{12}^{\rm SM} ( 1 + \kappa\, e^{i\sigma_q} ),\\
{\rm with\ }\kappa & \equiv & \left| \frac{M_{12}^{\rm LR}}{M_{12}^{\rm
      SM}}\right|,\\
\sigma_q & \equiv & {\rm arg}\, \frac{M_{12}^{\rm LR}}{M_{12}^{\rm
    SM}}\ =\ {\rm arg} \left( -\frac{V_{tb}^R V_{tq}^{R*}}{V_{tb}^L 
V_{tq}^{L*}} \right).\label{eq:def_sigma}
\end{eqnarray}
Note that the phase $\sigma_q$ is convention-independent and a
physical observable. The minus sign in the definition of $\sigma_q$
comes from the fact that, putting all CKM factors equal one,
$M_{12}^{\rm SM}$ and $M_{12}^{\rm LR}$ have different relative
sign. $\kappa$ is nearly independent of the flavour of the spectator
quark. Numerically, we find\footnote{Corrections in
  $(1-\eta^{LR}_2/\eta^{LR}_1)$ are smaller than neglected terms in $1/M_2^4$.}
\begin{equation}\label{eq:kappa}
\kappa = \frac{B_B^S(m_b)}{B_B(m_b)} \left[ \left(\frac{7\,{\rm
        TeV}}{M_H}\right)^2 + \eta_2^{LR}(m_b) \left( \frac{1.6\,{\rm
        TeV}}{M_2} \right)^2 \left\{0.051 - 0.013 \ln
\left(\frac{1.6\,{\rm
        TeV}}{M_2} \right)^2\right\}\right],
\end{equation}
which describes the full solution to within 5\% accuracy for $M_H>
7\,$TeV and $M_2>1.4\,$TeV. 

Let us now investigate the predictions of the SB--LR for and the
constraints from, respectively, the experimental data for $\Delta
m_B$. For the remainder of this section we consider $|M_{12}|$ as a
function of only two variables, $\kappa$ and $\beta$, instead of
expressing $\kappa$ in terms of $M_2$ and $M_H$. 
The mass difference in the $B_d^0$ system has been measured as
\cite{osc}
\begin{equation}\label{eq:delmBd}
\Delta m_{B_d} = (0.472\pm 0.016)\,{\rm ps}^{-1},
\end{equation}
whereas for the mass difference in the $B^0_s$ system, there exists only a
lower bound \cite{vancouver}:
\begin{equation}\label{eq:bound}
\Delta m_{B_s} > 12.4\,{\rm ps}^{-1};
\end{equation}
the SM expectation is \cite{fit}
\begin{equation}\label{eq:SMexp}
 \Delta m^{\rm SM}_{B_s} = (14.8\pm 2.6)\, {\rm ps}^{-1}.
\end{equation}
In the SM and with our values for the CKM angles, Eq.~(\ref{eq:CKMfix}), 
(\ref{eq:delmBd}) restricts the phase $\delta$ as
$\delta^{\rm SM} = 1.17\pm 0.44$ (taking into account that the
measured value of $\epsilon$ implies $\delta^{\rm SM}>0$), 
where the error comes mainly
from $f_B^2 \hat{B}_B$.
Taking into account the theoretical uncertainties on $f_B^2 \hat{B}_B$
and $m_t$, (\ref{eq:delmBd}) translates into
\begin{equation}\label{eq:5.25}
\left| (V_{tb}^L V_{td}^{L*})^2 (1+\kappa \, e^{i\sigma_d}) \right| =
  (6.7\pm 2.7)\cdot 10^{-5}.
\end{equation}
{}From this result we may derive a
constraint for $\kappa$: in the worst case of negative relative sign,
$\kappa$ evidently cannot be larger than
$$
\kappa^{\rm max} -1 = \frac{(6.7\pm 2.7)\cdot 10^{-5}}{
| (V_{tb}^L V_{td}^{L*})^2|},
$$
which roughly translates into
$$
\kappa < 3,
$$
which we will use in the following. For $B_s$ mixing, on the other
hand, we obtain from
(\ref{eq:bound}) the lower bound
\begin{equation}\label{eq:boundmBs}
\left|(V_{tb}^L V_{ts}^{L*})^2 (1 + \kappa e^{i\sigma_s})\right| > 9.6
\cdot 10^{-4}.
\end{equation} 
For the ratio of mass differences, the theory error is much smaller:
\begin{eqnarray}
\frac{\Delta m_{B_s}}{\Delta m_{B_d}} & = &
\frac{m_{B_s}}{m_{B_d}} \left( \frac{f_{B_s}}{f_{B_d}}\right)^2 
\frac{\hat{B}_{B_s}}{\hat{B}_{B_d}}\left| \frac{(V_{tb}^L
    V_{ts}^{L*})^2 (1+\kappa e^{i\sigma_s})}{(V_{tb}^L
    V_{td}^{L*})^2 (1+\kappa e^{i\sigma_d})}\right|\nonumber\\
& = &(1.31\pm 0.19) \left| \frac{(V_{tb}^L
    V_{ts}^{L*})^2 (1+\kappa e^{i\sigma_s})}{(V_{tb}^L
    V_{td}^{L*})^2 (1+\kappa e^{i\sigma_d})}\right|,\label{eq:aechz}
\end{eqnarray}
which translates into the bound
\begin{equation}\label{eq:boundmBs2}
\left| \frac{(V_{tb}^L
    V_{ts}^{L*})^2 (1+\kappa e^{i\sigma_s})}{(V_{tb}^L
    V_{td}^{L*})^2 (1+\kappa e^{i\sigma_d})}\right| > 17.2.
\end{equation}
\begin{figure}[p]
\vspace*{-1cm}
$$
\epsfysize=0.4\textheight
\epsffile{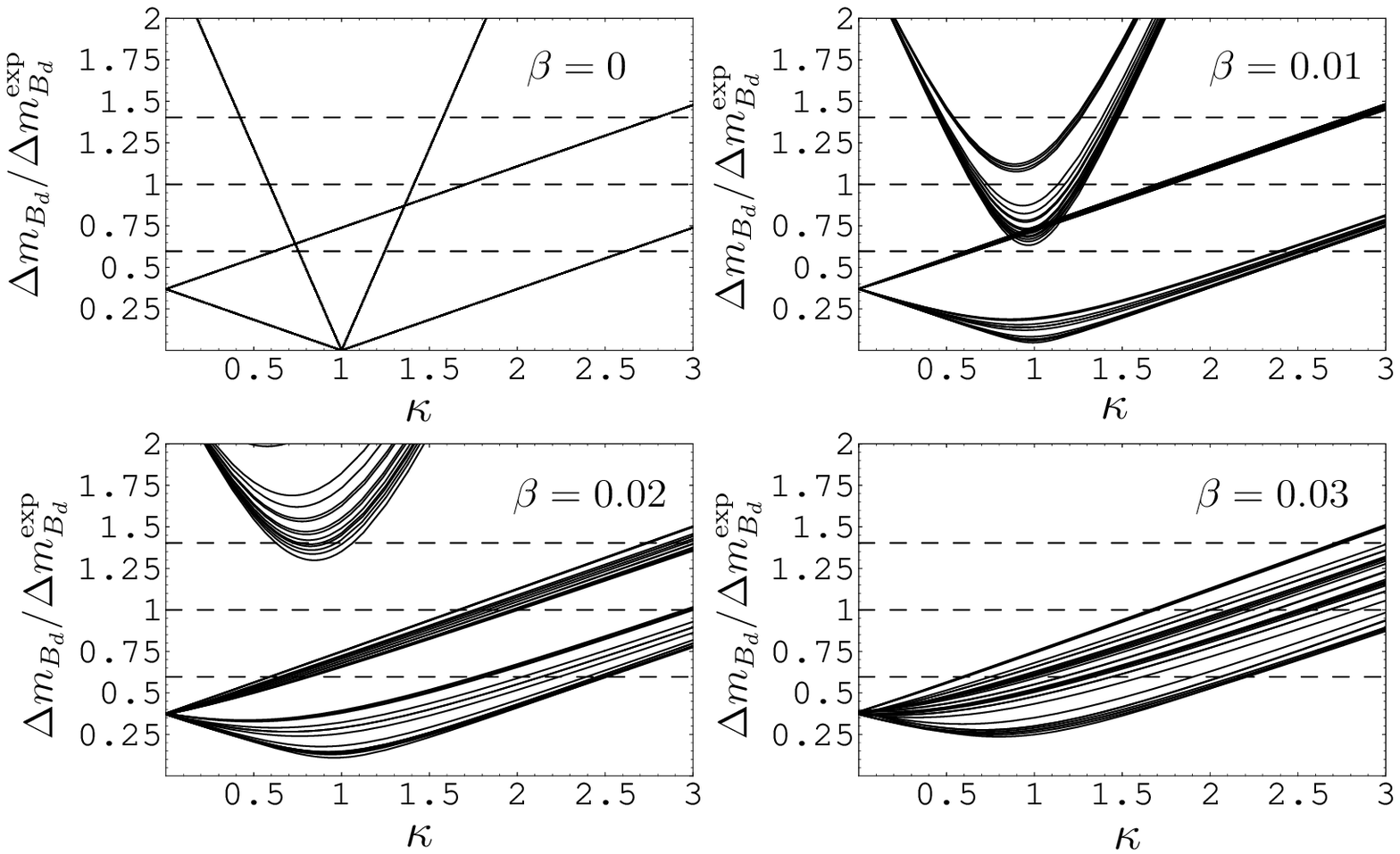}
$$
\vspace*{-0.5cm}
\caption[]{Constraints from $\Delta m_{B_d}$. Predictions of the SB--LR
 as functions of $\kappa$ for different values of $\beta$, 
$\beta=0,0.01,0.02,0.03$. Dashed
  lines are the experimental result and theory errors. 
The lower curves are class I solutions, the upper curves are class II.
}\label{fig:deltamB}
$$
\epsfysize=0.4\textheight
\epsffile{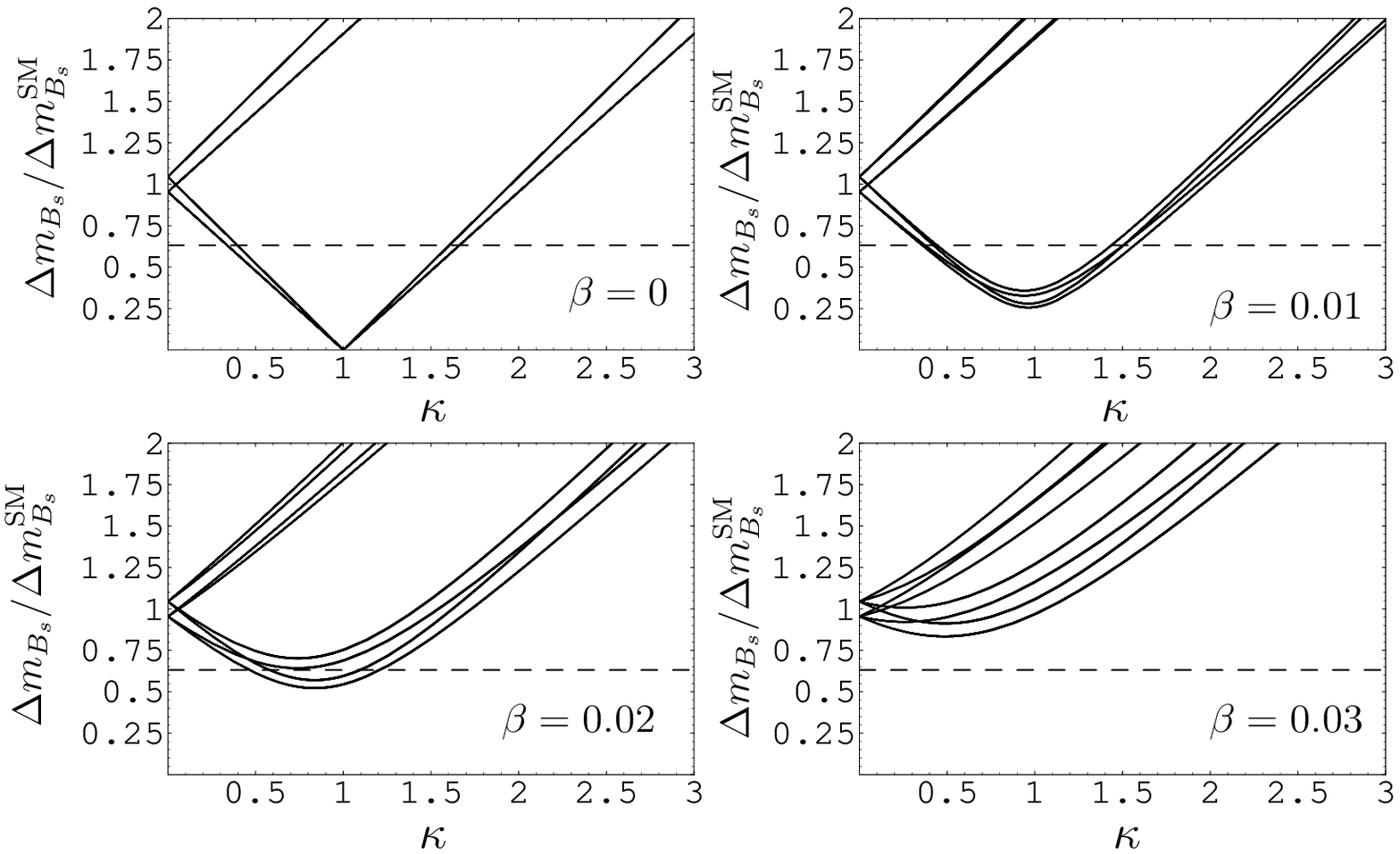}
$$
\vspace*{-0.5cm}
\caption[]{Like the previous figure, but for $\Delta
  m_{B_s}$. Normalization of vertical axis corresponds to the SM
  expectation $|V_{ts}| = 0.039$.
Dashed line is the lower bound on $\Delta m_{B_s}$.
}\label{fig:deltamBs}
\end{figure}
\begin{figure}[p]
\vspace*{-1cm}
$$
\epsfysize=0.4\textheight
\epsfbox{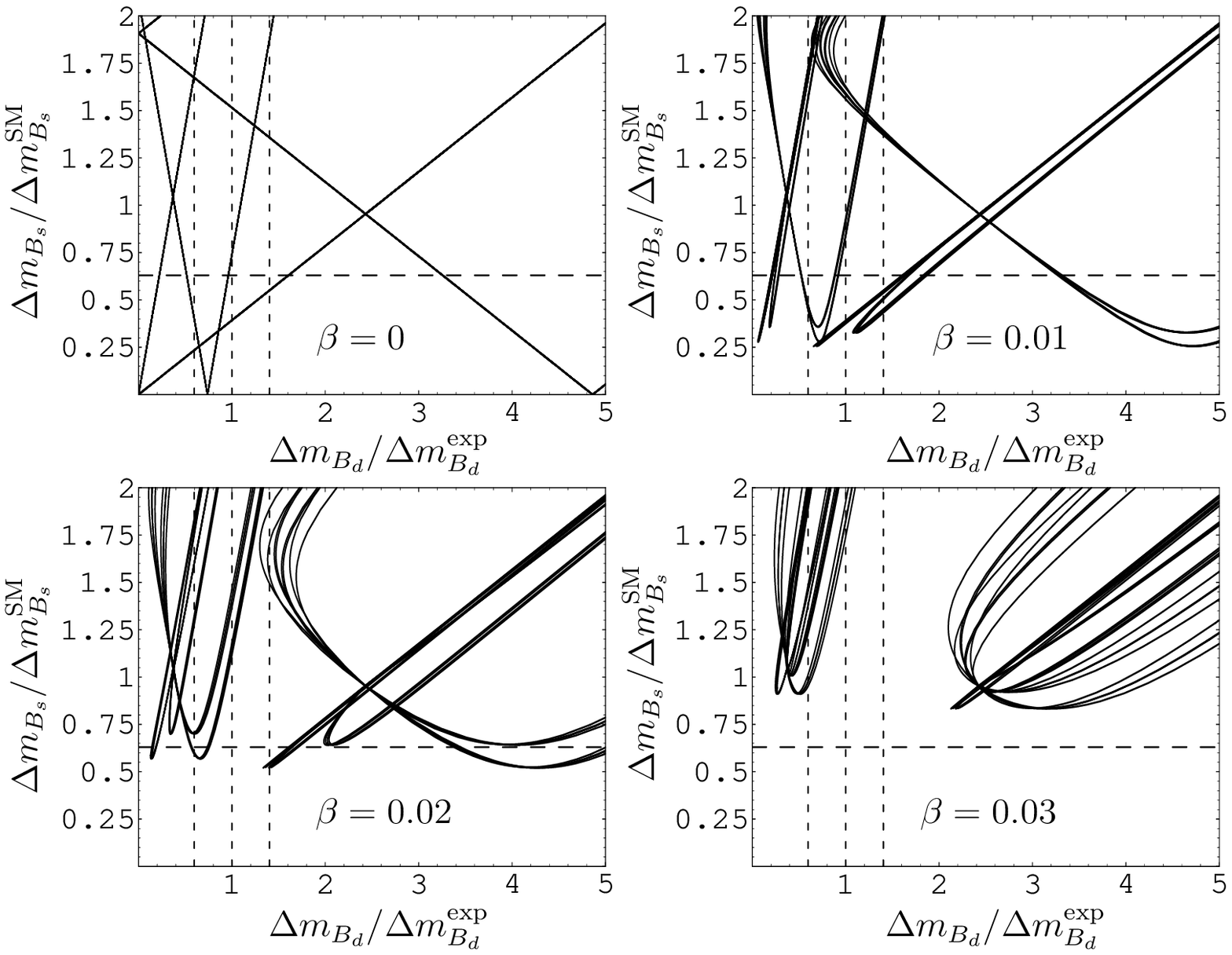}
$$
\vspace*{-30pt}
\caption[]{Correlation between predictions for $\Delta m_{B_d}$
  and $\Delta m_{B_s}$ for different
  values of $\beta$ as function of $\kappa$ in [0,3]. Short dashes denote
  experimental result and theory error for
  $\Delta m_{B_d}$, long dashes denote lower bound on
  $\Delta m_{B_s}$. The ensemble of lines in the left parts of the
  plots are class I solutions, the ones in the right
  halves are class II. 
}\label{fig:corrneu}
$$
\epsfysize=0.4\textheight
\epsfbox{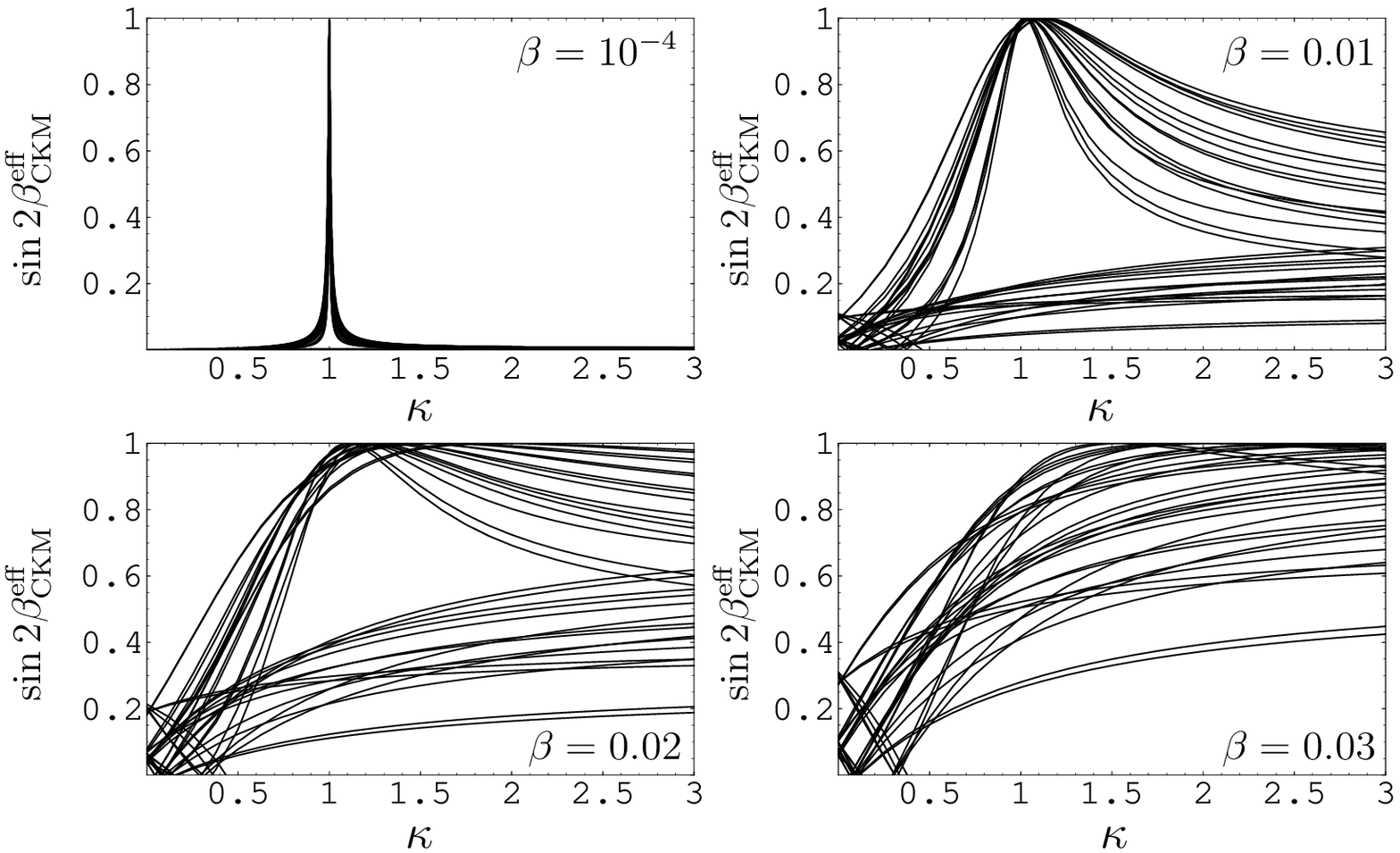}
$$
\vspace*{-30pt}
\caption[]{$\sin 2\beta^{\rm eff}_{\rm CKM}$ in the SB--LR as function of
  $\kappa$ for several values of $\beta$, $\beta =
  10^{-4}$, 0.01, 0.02, 0.03. 
Only solutions yielding positive values of $\sin
  2\beta^{\rm eff}_{\rm CKM}$ are shown.}\label{fig:7}
\end{figure}

In Fig.~\ref{fig:deltamB} we plot the left-hand side of
Eq.~(\ref{eq:5.25}), normalized by the central value on the right-hand side, as
function of $\kappa$ for several values of $\beta$. In
Fig.~\ref{fig:deltamBs} we plot
 $|(V^L_{ts} V_{tb}^{L*})^2(1+\kappa e^{i\sigma_s}|/0.039^2$, which is
 expected to be 1 in the SM. {}From the plots we conclude the
 following:
\begin{itemize}
\item the decoupling limit $\kappa\to 0$ of the SB--LR is
  excluded; this is due to the fact that the SM phase as extracted
  from $\Delta m_{B_d}$ is rather large, $\delta^{\rm SM}\sim 1$,
  whereas the SB--LR predicts values close to 0 or $\pi$; this result
  depends, however, on our specific choice of the CKM angles,
  Eq.~(\ref{eq:CKMfix}); yet, we
  shall derive the experimental exclusion of the decoupling limit
   unambiguously from the analysis of $\epsilon$
  in the next section;
\item class II solutions are excluded for $\beta\geq 0.021$;
\item class I solutions require $\kappa>0.52$, class II solutions $\kappa
  >0.42$. This means that the Higgs contributions to $\kappa$ are
  essential, see Eq.~(\ref{eq:kappa}). 
\end{itemize}
We next plot the correlations between predicted values for $\Delta
m_{B_s}$ and $\Delta m_{B_d}$, Fig.~\ref{fig:corrneu}. This plot, too,
illustrates the exclusion of the decoupling limit of the SB--LR, which
corresponds to the two crossing-points of the different classes of
solutions, best visible for $\beta = 0$. It is also evident
that the SB--LR can comfortably accomodate any non-standard value of $\Delta
m_{B_s}$ as well as the expected one, Eq.~(\ref{eq:SMexp}). 
There is
in particular a large fraction of class I and II solutions that
predict very large values of $\Delta m_{B_s}$, so that a measurement
of $\Delta m_{B_s}$ close to its SM expectation will effectively
constrain the parameter space of the SB--LR.

\subsection{Constraints from \protect{\goldd}}

Other interesting constraints can be obtained from the
measurement of the CP asymmetry in \goldd.
 Recently, the CDF collaboration has reported the
following result \cite{CDF}:
\begin{equation}\label{eq:CDF}
a_{\rm CP} = \frac{\Gamma(\bar B_d^0(t)\to J/\psi K_S^0) - 
\Gamma(B_d^0(t)\to J/\psi K_S^0)}{\Gamma(B_d^0(t)\to J/\psi K_S^0) +
\Gamma(\bar B_d^0(t)\to J/\psi K_S^0)} = (0.79^{+0.41}_{-0.44})\,
\sin\,(\Delta m_B t),
\end{equation}
and with 90\% probability $a_{\rm CP}/\sin\,(\Delta m_B t)>0$. 
In the SM, $a_{\rm CP}$ measures just $\sin 2\beta_{\rm CKM}$ with
\begin{equation}\label{eq:betaCKM}
\beta_{\rm CKM} = 
\arg \left( -\frac{V_{cd}^L V_{cb}^{L*}}{V_{td}^L V_{tb}^{L*}}\right);
\end{equation}
the SM model expectation is $\sin \,2\beta^{\rm SM}_{\rm CKM} 
= 0.73^{+0.05}_{-0.06}$ \cite{fit}.

In the SB--LR, however, we have to interpret the measurement differently. For
any B decay into a final CP eigenstate $f_{\rm CP}$, one can define a
convention and parametrization invariant quantity $\lambda$ by
$$
\lambda\equiv \left(\frac{q}{p}\right)_B \frac{\bar
  A_{f_{\rm CP}}}{A_{f_{\rm CP}}}
$$
with the amplitudes $A_{f_{\rm CP}} = A(B^0\to f)$ and $\bar A_{f_{\rm
    CP}} = A(\bar B^0\to f)$ and the mixing amplitude 
$(q/p)_B\simeq-\exp(-i\phi_M^B)$, defined
in (\ref{eq:phi}).
In the case of vanishing direct CP violation, the time-dependent CP
asymmetry can be written as
$$
a_{\rm CP} = {\rm Im}\, \lambda \; \sin (\Delta m_B t),
$$
so that we have
\begin{equation}
{\rm Im}\, \lambda(B^0 \to J/\psi K_S^0) = 0.79^{+0.41}_{-0.44}.
\end{equation}
The expression for $\lambda$ itself reads
$$
\lambda(B^0\to J/\psi K_S^0) = \exp(-i\phi_M^{B_d})\, \eta_{J/\psi K_S^0}\,
\left( \frac{V^L_{cb} V^{L*}_{cs}}{V^{L*}_{cb} V^L_{cs}}  \right)
\exp[i (-\phi_{CP}^B + \phi_{CP}^K)] \exp(i\phi_M^K).
$$
$J/\psi K_S^0$ is CP odd, hence $\eta_{J/\psi K_S^0}=-1$, and the K
mixing phase was defined in Eq.~(\ref{eq:phi}).
The imaginary part reads
\begin{equation}
\sin 2\beta^{\rm eff}_{\rm CKM} \equiv
{\rm Im}\, \lambda(B^0\to J/\psi K_S^0) = \sin \left[ 2 \beta_{\rm CKM}  +
\arg \left( 1 + \kappa e^{i \sigma_d}\right) -
 \arg \left( 1 + \frac{M_{12}^{\rm K,LR}}{M_{12}^{\rm K,SM}}\right)\right].
\end{equation}
The contribution from K mixing, the third term in
square brackets, is numerically small and can be neglected for
the moment.

In Fig.~\ref{fig:7} we plot $\sin 2\beta^{\rm eff}_{\rm CKM}$ as function of
$\kappa$ for several values of $\beta$, where we only show solutions
that yield $\sin 2\beta^{\rm eff}_{\rm CKM}>0$. It is obvious that the SM
expectation $\sin 2\beta^{\rm eff}_{\rm CKM}\approx 0.75$ can be
accomodated 
by a
number of solutions. It is also visible that for $\beta < 0.03$, there
are roughly two branches of solutions, one with small $\sin
2\beta^{\rm eff}_{\rm CKM}<0.4$, the other one spanning all possible values
between 0 and 1. A measurement of $\sin 2\beta^{\rm eff}_{\rm CKM}$ around its
SM expectation would favour either $\kappa \approx 0.6$ or
$\kappa>1.2$. Any more detailed analysis requires to take into account
the constraints from K mixing.

\section{Constraints from the K System }
\setcounter{equation}{0}

While the K system was the first one to be analysed in SB--LR, it remains 
plagued by theoretical uncertainties.\footnote{It should not be
  forgotten that the same is actually true for the SM.}
The main observables to be considered are obviously $\Delta m_K$, 
$\epsilon$ and $\epsilon'$. The formulas for $\Delta m_K$ are
analogous to those for $\Delta m_B$ discussed in the last
section. To be specific, we use the LO QCD corrected formulas for
$M_{12}^{\rm LR}$ of Ref.~\cite{EGNPB} with the bag-factor $B_K^S = 1$;
radiative corrections to $M_{12}^{\rm SM}$ are taken from
Ref.~\cite{buhmann}; we also use $\hat{B}_K = 0.89$ \cite{laurent}.
As for $\epsilon$, in the SM, it 
is usually written as
\begin{equation}\label{eq:grummel}
\epsilon = \frac{1}{\sqrt{2}}\, e^{i\pi/4}\left( \frac{{\rm Im}\, 
M_{12}}{\Delta  m_K} + \xi_0\right)
\end{equation}
with 
\begin{equation}
\xi_0  =  \frac{{\rm Im}\,a_0}{{\rm Re}\,a_0},\qquad
a_0^*  =  \langle \pi\pi{\rm(I=0)} | -i{\cal H}_{\rm eff}^{|\Delta
  S|=1}| \bar K^0\rangle_{\rm weak},\label{eq:a0}
\end{equation}
where the matrix element in (\ref{eq:a0}) does contain only the weak
phase, but no strong final-state rescattering phases. As the
derivation of this formula includes some relations and approximations
that need not be valid in the SB--LR, we rederive it, following
the transparent discussion given in Ref.~\cite{nakada}. 

The parameter $\epsilon$ measures essentially the phase-difference
between $M_{12}$ and $\Gamma_{12}$, where $\Gamma_{12}$ is the matrix
element of the decay matrix $\Gamma$ over the $K^0$ and $\bar K^0$
states. Introducing
$$
\delta \theta_{M/\Gamma} = \arg M_{12} - \arg \Gamma_{12},
$$
one finds for the hypothetical case of no CP violation
$$
\delta \theta_{M/\Gamma} = \pi,
$$
which essentially follows from the fact that $\Delta m_K \equiv m_L - m_s
>0$, but $-\Delta\Gamma_K \equiv \Gamma_L-\Gamma_S <0$. Note that only the
phase-difference $\delta \theta_{M/\Gamma}$ is an observable and
convention-independent quantity, but not $\arg M_{12}$ and $\arg
\Gamma_{12}$ separately, which depend on the K analogue of
the arbitrary
phase $\phi_{CP}$, introduced in (\ref{eq:Bfac}) for B's, and on the
$s$ and $d$ quark phases, i.e.\ on the parametrization of
the quark mixing matrices. For the
analysis of $\epsilon$, it proves convenient to choose
$\phi_{CP}=\pi$, which we shall use in the remainder of this
section. One then has --- still for the case of no CP violation and
using the Maiani parametrization of the CKM matrix --- $\arg
M_{12} = 0$ and $\arg\Gamma_{12} = \pi$. In the real CP violating
world, $\delta \theta_{M/\Gamma}$ is only slightly different from
$\pi$: making use of the fact that
 K decays are dominated
by the  $2\pi$ channel with isospin 0 (the famous $\Delta I=1/2$ rule), a 
measure of CP violation in the interference of mixing and decay, i.e.\
of $\delta
\theta_{M/\Gamma}$, is given by $\epsilon$, which is
defined in such a way as to contain no effects from direct
CP violation and which can be written as \cite{nakada}:
\begin{eqnarray}
\epsilon & = & -\frac{x}{4 x^2+1} \left\{1+(2 x) i\right\}\, \sin 
\delta\theta_{M/\Gamma},\\
{\rm with\ }x & = & \frac{\Delta m_K}{\Delta \Gamma_K} \ =\ 0.478\pm 0.002 \approx
0.5,\nonumber
\end{eqnarray}
so that
\begin{equation}\label{eq:bis_zum_bitteren_ende}
\epsilon \approx \frac{1}{2\sqrt{2}}\, e^{i\pi/4}\,(- \sin
\delta\theta_{M/\Gamma}).
\end{equation}
$\Gamma_{12}$ can, in contrast to $M_{12}$, not be calculated
accurately from theory, and one exploits the dominance of decays into the
$2\pi({\rm I=0})$ final state to derive\footnote{The factor $e^{i\pi/2}$
  comes from the fact that $\arg\Gamma_{12}$ is related to
  the matrix element of ${\cal H}_{\rm eff}$, whereas we have defined
  $a_0^*$ as matrix element of $-i{\cal H}_{\rm eff}$.}
\begin{equation}\label{eq:spiel_mir_das_lied_vom_tod}
\arg \Gamma_{12} \approx - 2 \arg \left( a_0 e^{i\pi/2}\right).
\end{equation}
Combining Eqs.\ (\ref{eq:bis_zum_bitteren_ende}) and
(\ref{eq:spiel_mir_das_lied_vom_tod}), we finally obtain
\begin{equation}\label{eq:finally!}
\epsilon = \frac{1}{2\sqrt{2}}\, e^{i\pi/4}\sin\left( \arg M_{12} + 2
  \arg a_0\right).
\end{equation}
In contrast to Eq.~(\ref{eq:grummel}), this formula also allows to demonstrate
explicitly that $\epsilon$ does not depend on the parametrization of
the quark mixing matrices: $M_{12}$ contains the generic CKM factor
$(\lambda_i^{AB})^2$, $A,B\in \{ L,R\}$, $i\in \{u,c,t\}$, whereas
$a_0$ contains the factor $\lambda^{AB*}_i$, so that
phase-redefinitions of $V_L$ and $V_R$ cancel in the sum
(\ref{eq:finally!}). 

The experimental result for $\epsilon$, $|\epsilon| = (2.280\pm 0.013)\cdot
10^{-3}$ \cite{PDG}, implies
\begin{equation}\label{eq:stick_to}
\arg M_{12} + 2 \arg a_0 = (6.449 \pm 0.037)\cdot  10^{-3}.
\end{equation}
In the SM, both terms in the above sum are small so that arg can be
replaced by the ratio of imaginary to real part. In addition, as $\arg
M_{12} \approx 0$ in our phase-convention, one also has ${\rm
  Re}\,M_{12}\approx |M_{12}|$, which is just $\Delta m_K/2$, so that
one can approximate (\ref{eq:finally!}) by (\ref{eq:grummel}) to
excellent accuracy. As $2\arg a_0\ll \arg M_{12}$, this term can
safely be neglected in the SM; in the SB--LR this is no
longer true: the contributions of $W_R$ to $a_0^{\rm LR}$ have been calculated 
in Refs.~\cite{EGNPB,JMF}, but involve considerable theoretical
uncertainties. {}From \cite{EGNPB,JMF}, we find
$$
2 |\arg a_0^{\rm LR}| < 0.005\cdot \left(\frac{1\,{\rm TeV}}{M_2}\right)^2.
$$
In view of the theoretical uncertainty of $a_0^{\rm LR}$, which
involves a number of only poorly known hadronic matrix elements,
we prefer to include it into the uncertainty of $\arg M_{12}$ in
(\ref{eq:stick_to}). Assuming that the unknown Higgs contributions are
not larger than those from $W_R$, we include twice the value of
$2 |\arg a_0^{\rm LR}|$ in the uncertainty of $\arg M_{12}$ and thus
find the constraint
\begin{eqnarray}\label{eq:wers_glaubt_wird_selig}
6.375\cdot 10^{-3}-0.01 \cdot \left(\frac{1\,{\rm TeV}}{M_2}\right)^2
  & < & \widetilde{\theta}_{M} < 6.523\cdot 10^{-3} + 
0.01 \cdot \left(\frac{1\,{\rm TeV}}{M_2}\right)^2\\
{\rm with\ } \widetilde{\theta}_{M} & = & 
\left|\frac{2\, {\rm Re}\,M_{12}}{\Delta m_K}\right| \arg M_{12},\nonumber
\end{eqnarray}
which also takes into account 2 experimental standard deviations and
where we have rescaled Re$\,M_{12}$ to its experimental value.
\begin{figure}
$$
\epsffile{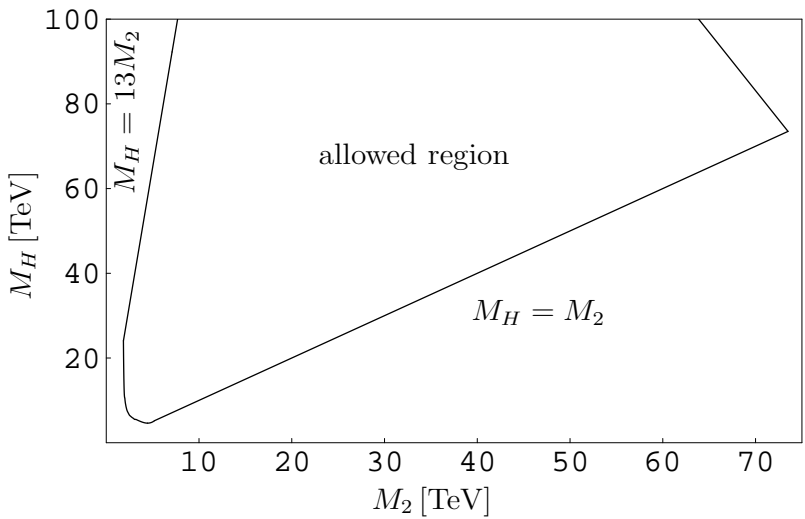}
$$
\caption[]{Allowed values for $M_2$ and $M_H$ from the K physics constraints.
}\label{fig:yet_another_figure}
\end{figure}

Finally, the expression for $\epsilon'$ reads
$$
\epsilon' \approx \frac{1}{\sqrt{2}}\, \exp^{i \pi/4}\,
    \frac{{\rm Re}\, a_2}{{\rm Re}\,
  a_0} \,(\xi_2-\xi_0),
$$
where $\xi_2$ is defined in analogy to $\xi_0$ for the $2\pi{\rm
  (I=2)}$ final state. In view of the large theoretical uncertainties
associated with the precise value of ${\rm Re}(\epsilon'/\epsilon)$,
we only require the SB--LR to predict a positive value.

Let us now discuss the constraints to be obtained from the three
observables $\Delta m_K$, $\epsilon$ and $\epsilon'$.
First, we consider the decoupling limit $M_2,M_H\to\infty$. In this
limit we find
$$
\widetilde{\theta}_{M} < 2.9\cdot 10^{-3},
$$
which is less than half of the experimental value and is related to
the smallness of the standard CKM phase $\delta$ in the SB--LR. {}From
this result, which is insensitive to the exact value of the uncertain
CKM angle $\theta_{13}$, we firmly conclude that
the {\em decoupling limit $M_2,M_H\to\infty$ is experimentally excluded}.

We next investigate the allowed region in the space of mass-parameters
imposed by the constraint (\ref{eq:wers_glaubt_wird_selig}) and the
one on $\Delta m_K$, Eq.~(\ref{eq:wuerg}). The result is plotted in
Fig.~\ref{fig:yet_another_figure}. 
We find in particular the following lower bounds on the extra boson
masses:
\begin{equation}\label{eq:idiotie1}
M_2 > 1.85\,{\rm TeV},\qquad M_H > 5.2\,{\rm TeV}.
\end{equation}
The bound for $M_2$ is in the ballpark of the usually obtained values,
cf.\ Ref.~\cite{EGNPB}, the one on $M_H$ is smaller, the reason being
that, in contrast to all previous analyses, we did not assume charm quark
dominance for $M_{12}^H$, but also included the top
quark contributions which can
destructively interfere with the charm quark ones, thus lowering the
limit on $M_H$. The experimental limit on $\widetilde{\theta}_M$,
i.e.\ $\epsilon$, implies also (not very constraining) 
upper bounds on the extra boson masses:
\begin{equation}\label{eq:idiotie2}
M_2 < 73.5\, {\rm TeV},\qquad M_H < 230\,{\rm TeV}.
\end{equation}
We recall that all these limits and bounds are to be modified by the
inclusion of B physics constraints.
This concludes our discussion of constraints
from K physics.

\section{Combining Constraints from K and B System}
\setcounter{equation}{0}

\begin{figure}[t]
$$
\epsffile{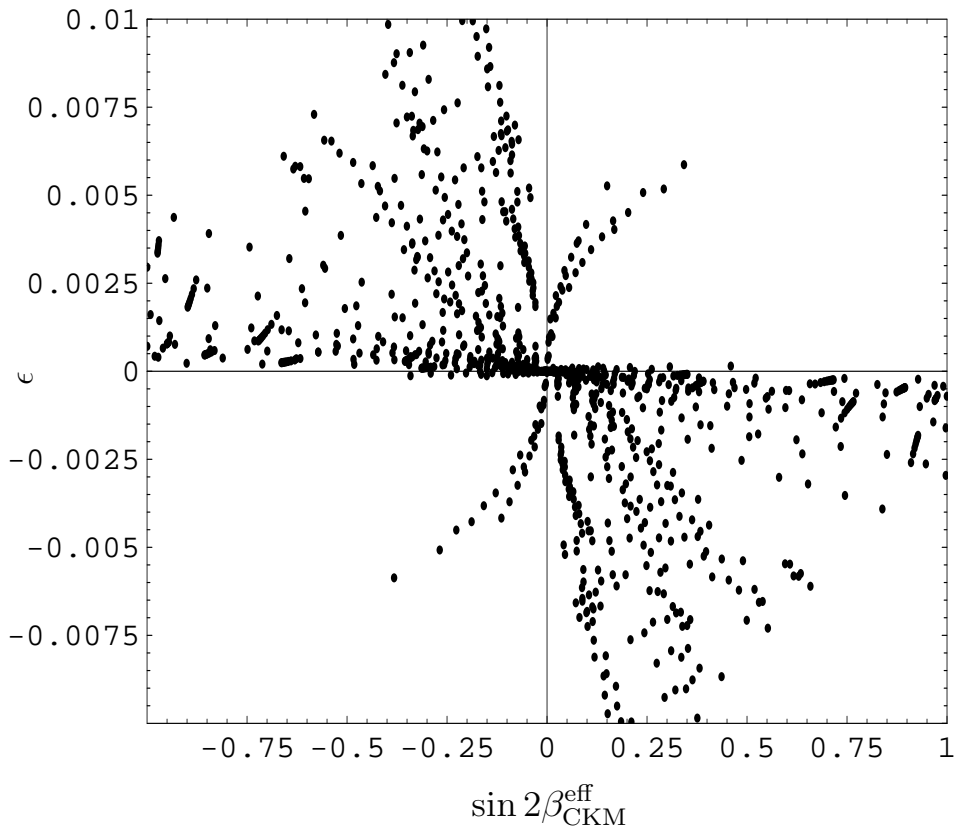}
$$
\caption[]{Allowed values for the CP violating parameters $\epsilon$
  and $\sin 2\beta_{\rm CKM}^{\rm eff}$ whose input parameters
 pass the $\Delta m$ cuts and yield the correct sign of
 Re\,$\epsilon'$.}\label{fig:sinepscorr}
\end{figure}

\begin{figure}
$$\epsffile{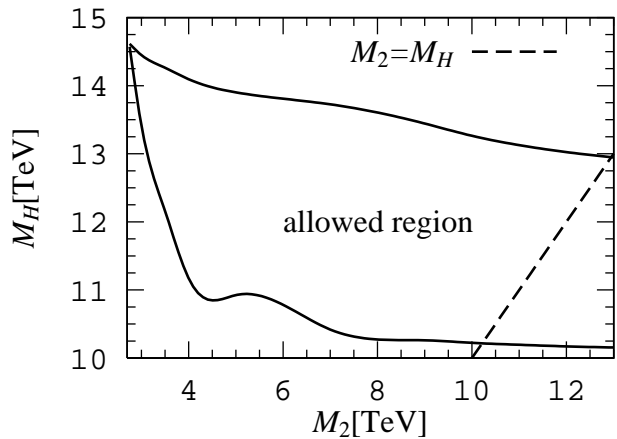}$$
\vspace*{-35pt}
\caption[]{Allowed region in $(M_2,M_H)$, taking into account all
constraints.}\label{fig:the_final_one_I_swear}
\end{figure}

Combining all the constraints from $\Delta m_K$, $\Delta m_{B_d}$,
$\Delta m_{B_s}$, $\epsilon$, $\epsilon'$ and $\sin 2\beta_{\rm
  CKM}^{\rm eff}$, our main finding is that, although the values of
the CP
conserving observables can be reproduced by a large range of input parameters,
this is not the case for the CP violating ones: the
crucial point is a strong anti-correlation between the signs of
Re$\,\epsilon$ and $\sin 2\beta_{\rm CKM}^{\rm eff}$, which are both
known to be positive from experiment. We illustrate this point in
Fig.~\ref{fig:sinepscorr}, where we plot the values of $\epsilon$ (to
be precise: $\epsilon \cdot e^{-i\pi/4}$) vs.\
$\sin 2\beta_{\rm CKM}^{\rm eff}$ for all sets of input parameters
($n$, $\beta$, $M_2$, $M_H$) with $2\,{\rm TeV} \leq M_2\leq 50\,
{\rm TeV}$ and $M_2 \leq M_H\leq 13 M_2$ that pass the cuts on the
mass differences (with a 50\% uncertainty on $\Delta m_{B_d}$ to
account for the uncertainty of the CKM angles) and on the sign of Re$\,
\epsilon'$. It is obvious that only a few sets of input parameters can
reproduce the observed sign of both $\epsilon$ and $\sin 2\beta_{\rm
  CKM}^{\rm eff}$. We find that the class I quark mass signature no.\
31 is the only one to accomplish that, and thus the only one of the
initial 64 signatures to
survive all cuts. A closer inspection shows that the maximum
possible $\sin 2\beta_{\rm CKM}^{\rm eff}$ correlated with
$\widetilde{\theta}_M$ in the range given in
(\ref{eq:wers_glaubt_wird_selig}) is
$$
\sin 2\beta_{\rm CKM}^{\rm eff,max} = 0.1,
$$
which is incompatible with the SM expectation $\sin 2\beta_{\rm CKM}^{\rm
  SM}\approx 0.75$.
The exclusion of all quark mass signatures except for one also cuts
deeply into the allowed range for $M_2$ and $M_H$. For fixed $M_2$
(and $\beta$), we have the following constraints on $M_H$:
\begin{itemize}
\item a lower bound from $\Delta m_K^{\rm LR} < \Delta
  m_K^{\rm exp}$;
\item an upper bound from $\sin 2\beta_{\rm SM}^{\rm eff}>0$ (because
  $\sin 2\beta_{\rm SM}^{\rm eff}<0$ for $M_H\to\infty$);
\item a lower bound from the upper limit on $\widetilde{\theta}_M$,
  Eq.~(\ref{eq:wers_glaubt_wird_selig});
\item an upper bound from the lower limit on $\widetilde{\theta}_M$.
\end{itemize}
The allowed region in $(M_2,M_H)$ (also taking into account the
constraints from $\Delta m_B$) thus gets very much restricted, as
shown in Fig.~\ref{fig:the_final_one_I_swear}. 
We find the bounds
$$
2.75\,{\rm TeV} < M_2 < 13\,{\rm TeV},\qquad 10.2\,{\rm TeV} < M_H <
14.6\,{\rm TeV},
$$
which improve considerably on those from K physics alone,
Eqs.~(\ref{eq:idiotie1}) and (\ref{eq:idiotie2}). The predictions for
$\Delta m_{B_s}$ are in the range $(0.6-1.1)\,\Delta m_{B_s}^{\rm
  SM,exp}$, i.e.\ a measurement of $\Delta m_{B_s}$ close to its SM
expectation would not pose any additional constraint.

As for $\beta$, we find that 
$$
 0.009 < \beta < 0.0327,
$$
i.e.\ the combination of K and B constraints only results into a lower
bound on $\beta$, but does not improve the maximum one, which is still
given by (\ref{eq:ra}). 

\section{Summary and Discussion}
\setcounter{equation}{0}

In this paper we have investigated in detail the present status of the
left-right symmetrical model with spontaneous CP violation, based on the
gauge group \model\/. The parameter space of this model 
includes the masses of the
predominantly right-handed charged gauge boson, $M_2$, of FC neutral and
charged Higgs bosons, which we have assumed to be degenerate with a common
mass $M_H$, as well as the parameter $\beta$, measuring the size of the VEV
of the Higgs bidoublet $\Phi$, which characterizes the spontaneous
breakdown of CP symmetry, and $n$, the 64 different quark mass
signatures, which are observable in the SB--LR.
In contrast to previous publications, e.g.\
\cite{EGNPB,langacker,prades,bloedsinn}, in which the constraints on
the model from K and B physics were treated separately, ours is the
first one to consider them in a coherent way and using the exact
results for the CKM phases instead of the small phase approximation. 
We have concentrated on
experimental constraints imposed by the mass differences $\Delta
m_{K,B}$ and observables describing CP violation, i.e.\ $\epsilon$,
$\epsilon'$ and $a_{\rm CP}(B\to J/\psi K_S)$. In view of large
theoretical uncertainties, we only use the sign, but not the absolute value of
Re$\,(\epsilon'/\epsilon)$ as a constraint, and we do not use the
electric dipole moment of the neutron. Our main finding is
that, although the K and B constraints can be met {\em separately} by
a large range of input parameters, it is their {\em combination} that
restricts the model severely. 
We find in particular
that the CP violating observables $\epsilon$ and $\sin 2\beta_{\rm
  CKM}^{\rm eff}$ are crucial: the sets of input parameters that pass
the constraints imposed by the meson mass differences $\Delta m_{K,B}$
yield to a large majority {\em opposite signs} of $\epsilon$ and 
$\sin 2\beta_{\rm  CKM}^{\rm eff}$. The combination of all constraints
yields the following results:
\begin{itemize}
\item all but one quark mass signatures are excluded, only class I
  solution no.\ 31 survives;
\item $a_{\rm CP}(B\to J/\psi K_S)\equiv \sin 2 \beta_{\rm CKM}^{\rm
    eff} < 0.1$, which is compatible with the present experimental
  result (\ref{eq:CDF}), but incompatible with the SM expectation
  0.75;
\item predictions for $\Delta m_{B_s}$ are in the range
  $(0.6-1.1)\,\Delta m_{B_s}^{\rm SM,exp}$;
\item the masses of the extra bosons are restricted to
$$
2.75\,{\rm TeV} < M_2 < 13\,{\rm TeV},\qquad 10.2\,{\rm TeV} < M_H <
14.6\,{\rm TeV};
$$
\item the value of $\beta$ is restricted to $0.009<\beta<0.0327$.
\end{itemize}

We would like to stress that our findings are largely independent
of the details of the scalar potential: the relevant neutral Higgs vertices
can be obtained essentially from the requirement of gauge-invariance of S
matrix elements, as discussed in Ref.~\cite{soni}. 
This does not apply to the charged Higgs vertices, which
in principle {\em do} depend on the specifics of the scalar potential:
 we thus have imposed the condition $M_H>M_2$ in order to suppress all
 contributions from charged Higgses (in particular box-diagrams).

We also would like to stress that our study does not claim to be 
exhaustive as we did not allow
the most crucial SM input parameters, the CKM angles and quark masses,
to float within their presently allowed ranges. Taking into account
these uncertainties would certainly affect
the phases of the CKM matrices and thus mainly
show up in the CP violating observables, which, as we have shown, are
crucial. It is thus not to be excluded that an analysis of the
input parameter uncertainties
would result in increasing the viable LR parameter ranges,
but we doubt that it will change the anticorrelation
between the signs of $\epsilon$ and $\sin 2\beta_{\rm CKM}^{\rm eff}$,
which implies a small maximum value of
$\sin 2\beta_{\rm CKM}^{\rm  eff}$  attainable in the model.

Another limitation of the present analysis is that we have kept the
ratio of Higgs VEVs, $r$, constant and equal to $m_b/m_t$.
As stressed before, this quantity governs the amount of mixing between L
and R bosons. For most observables, the relevant parameter is 
$\beta \sim 2r \sin\alpha$, on which our analysis is based.
Reducing the value of $r$ while keeping $\beta$ fixed 
(remember that $\tan[\beta/2]\leq m_b/m_t$) would not affect our conclusions.
Increasing $r$, however, has an impact on the imaginary parts of $a_0$
and $a_2$, and from there on the values of $\epsilon$ and
$\epsilon'$. Such an increase is strongly disfavoured if we take into
account the constraint from the neutron's EDM. Our most important
result, namely the bound on $\sin2 \beta_{\rm CKM}^{\rm eff}$ 
%from B -Bbar mixing however 
is, however, not affected by these considerations.

\section*{Acknowledgements}

P.B.\ is supported by DFG through a Heisenberg fellowship. 
 J.M.F.\ gratefully acknowledges 
 hospitality and financial support
from the Theory Division of CERN, where this work got started.
He is also supported in part by the ``Actions de Recherche
Concert\'{e}es'' of the
``Direction de la Recherche Scientifique --- Communaut\'{e} Fran\c{c}aise de
Belgique'', IISN--Belgium, convention no.\ 4.4505.86. J.M.\
 acknowledges financial support from a Marie Curie EC
Grant (TMR--ERBFMBICT 972147).

\appendix
\renewcommand{\theequation}{\Alph{section}.\arabic{equation}}

\section*{Appendix}
\section{Non-factorizable Contributions to $B_B^S/B_B$}
\setcounter{equation}{0}

We estimate the ratio of bag-factors $B_B^S/B_B$, which enters the
matrix element $M_{12}$ describing $B^0$--$\bar{B}^0$ mixing, by two
methods, the $1/N_c$ expansion and QCD sum rules. Some aspects of the
$1/N_c$ expansion for $B_B$ have also been
discussed in \cite{bardeen}.

To leading order in the $1/N_c$ expansion, it follows from
(\ref{eq:Bfac}) that $B_B = 3/4$ and from (\ref{eq:OS}) that $B_B^S =
1/(1+m_b^2/(6 m_B^2))$, so that
\begin{equation}\label{eq:A1}
\frac{B_B^S}{B_B} \stackrel{N_c\to\infty}{=} \frac{4}{3}\,
\frac{1}{1+\frac{1}{6}\, \frac{m_b^2}{m_B^2}} \approx 1.2.
\end{equation}
As factorization is exact in the large $N_c$ limit, $B_B$ becomes
scale-independent and it is not clear at what scale (\ref{eq:A1}) is
valid. 
Another approach more suited to include the scale-dependence 
is provided by QCD sum rules \cite{SVZ}. QCD sum
rules for $B_B$ have already been discussed in \cite{QCDBB}; our
results for $B_B^S$ are new. We consider the correlation functions
\begin{eqnarray*}
\Pi(p^2,p'^2) & = & i^2\!\!\int\!\! d^4x d^4y e^{i(px-p'y)} \langle 0
| T j_B^\dagger(x) O^{SM}(0) j_B^\dagger(y)| 0 \rangle,\\
\Pi^S(p^2,p'^2) & = & i^2\!\!\int\!\! d^4x d^4y e^{i(px-p'y)} \langle 0
| T j_B^\dagger(x) O^{S}(0) j_B^\dagger(y)| 0 \rangle,
\end{eqnarray*}
with the currents and operators
\begin{eqnarray*}
j_B & = & (m_b+m_d) \bar d i\gamma_5 b,\\
O^{SM} & = & (\bar d b)_{V-A}\,(\bar d b)_{V-A},\\
O^S & = & (\bar d b)_{S-P}\,(\bar d b)_{S+P}.
\end{eqnarray*}
In the standard QCD sum rules philosophy, $\Pi$ and $\Pi^S$ are on the
one hand
calculated in a local operator product expansion in the deep Euclidean
region $p^2,p'^2\ll 0$ and on the other hand analytically continued to
the Minkowskian region by dispersion relations, saturated by the
hadronic ground state. Equating these two representations yields ---
after some technicalities which are well known to the experts and
which we refrain from describing in this appendix --- QCD sum rules
for $B_B$ and $B_B^S$, respectively. It turns out that the leading
non--factorizable contributions come from the dimension 5 mixed
condensate $\langle \bar d \sigma g G d\rangle$, which is enhanced by
a factor $m_b$; the sum rules read:
\begin{eqnarray*}
\frac{m_B^4 f_B^2}{(p^2-m_B^2)(p'^2-m_B^2)}\,\frac{8}{3}\, f_B^2 m_B^2
B_B & = & \Pi_{\rm fact} + \Pi^{\langle 5\rangle} + \dots,\\
\frac{m_B^4 f_B^2}{(p^2-m_B^2)(p'^2-m_B^2)}\,2 f_B^2 m_B^2
B_B^S & = & \Pi_{\rm fact}^S + \Pi^{S,\langle 5\rangle} + \dots
\end{eqnarray*}
Here the dots denote subleading non-factorizable contributions from
$O(\alpha_s)$ perturbation theory and the gluon, four-quark and higher
condensates, which we neglect in our estimate.

We can now subtract the factorizable parts, perform
Borel-transformation and continuum subtraction (some of the
technicalities mentioned above), and with $B \equiv 1 + \Delta
B$ we obtain
\begin{equation}\label{eq:A2}
\frac{\Delta B_B^S}{\Delta B_B} = \frac{4}{3} \,
\frac{1}{\frac{m_B^2}{m_b^2} + \frac{1}{6}}\, \frac{\hat{B}
  \Pi^{S,\langle 5\rangle}_{\rm non-fact}}{\hat{B}
  \Pi^{\langle 5\rangle}_{\rm non-fact}}\,.
\end{equation}
Unfortunately, both $\Delta B$ in the numerator and denominator are
scale-dependent so that (\ref{eq:A2}) suffers from a large
scale-uncertainty. Following Ref.~\cite{bestseller}, we thus introduce
the LO RG-invariant quantities ($M^2$ is the Borel-parameter)
\begin{eqnarray*}
\hat{\Pi}(M^2) & = & \left[ \alpha_s(M^2/m_b)
\right]^{-\gamma_B/(2\beta_0)} \hat{B} \Pi(M^2),\\
\hat{\Pi}^S(M^2) & = & \left[ \alpha_s(M^2/m_b)
\right]^{-\gamma_{B^S}/(2\beta_0)} \hat{B} \Pi^S(M^2),
\end{eqnarray*}
where we have in particular taken into account that the natural scale
of QCD sum rules for heavy hadrons is $\mu = M^2/m_b$, see the
discussion in \cite{bestseller}. This allows us to calculate $\Delta
B_B^S/\Delta B_B$ directly at the scale $\mu = m_b$ with
\begin{eqnarray}
\frac{\Delta B_B^S(m_b)}{\Delta B_B(m_b)} & = & \frac{4}{3}\,
\frac{1}{\frac{m_B^2}{\bar{m}_b(m_b)^2} + \frac{1}{6}} \left[
\frac{\alpha_s(m_b)}{\alpha_s(M^2/m_b)}\right]^{(\gamma_{B^S}-\gamma_B)/
(2\beta_0)}\nonumber\\
& & \times \left( -\frac{1}{4} + \frac{1}{2}\,
\frac{\int_{m_b^2}^{s_0}  ds\,\frac{(s+m_b^2)(m_b^2-s)}{s^2}\,
  e^{-s/M^2}}{ \int_{m_b^2}^{s_0}  ds\,\frac{(s+m_b^2)m_b^2}{s^2}\,
  e^{-s/M^2}}\right)
\end{eqnarray}
with $\beta_0 = 11-2/3 \cdot 4$, $\gamma_{B^S} = -8$ and $\gamma_B =
2$; $s_0\approx 34\,$GeV$^2$ is the continuum threshold. 

\begin{figure}
$$
\epsffile{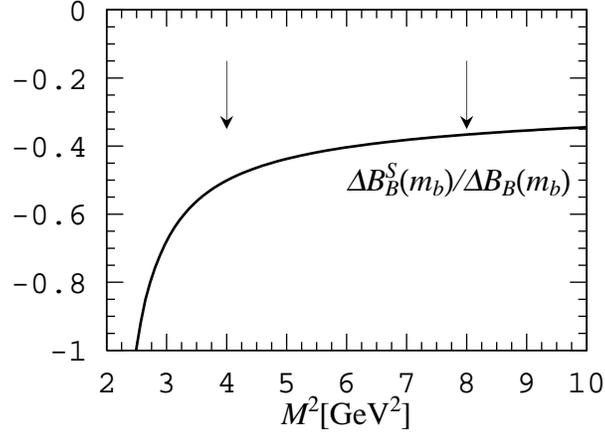}
$$
\vspace*{-30pt}
\caption[]{QCD sum rule for $\Delta
B_B^S(m_b)/\Delta B_B(m_b)$ as function of the Borel-parameter $M^2$;
input parameters: one-loop pole mass $m_b = 4.8\,$GeV, $s_0 =
34\,$GeV$^2$. The arrows denote the fiducial region for the 
Borel-parameter.}\label{fig:app}
\end{figure}
In Fig.~\ref{fig:app} we plot the sum rule as function of $M^2$; the
dependence on the $M^2$ as well as on $s_0$ and $m_b$ is very mild; we
find
\begin{equation}
\frac{\Delta B_B^S(m_b)}{\Delta B_B(m_b)} = -(0.43\pm 0.07),
\end{equation}
which, with $\Delta B_B(m_b) = -(0.1\pm 0.1)$ from lattice
calculations \cite{laurent}, yields
\begin{equation}
\frac{B_B^S(m_b)}{B_B(m_b)} = 1.16\pm 0.14.
\end{equation}
This result agrees perfectly well with that from $1/N_c$
expansion, Eq.~(\ref{eq:A1}). We thus quote as our final result
\begin{equation}
\frac{B_B^S(m_b)}{B_B(m_b)} = 1.2\pm0.2.
\end{equation}


\begin{thebibliography}{99}

\bibitem{houart} J.-M.~Fr\`ere et al.,
Phys. Lett. B {\bf 314} (1993) 289. % (hep--ph/9301228).

\bibitem{liu}   J.-M.\ Fr\`{e}re  and J. Liu, Nucl.\ Phys.\   {\bf
B324}
  (1989) 333.

\bibitem{langacker} P.~Langacker and S.U.\ Sankar,
Phys. Rev. D {\bf 40} (1989) 1569.

\bibitem{EGNPB} G. Ecker and W. Grimus, Nucl.\ Phys.\ {\bf B258}
  (1985) 328.

\bibitem{fit} F. Parodi, P. Roudeau and A. Stocchi, Preprint hep--ex/9903063.

\bibitem{prades}
G.~Barenboim, J. Bernabeu and M. Raidal,
Nucl.\ Phys.\ {\bf B478} (1996) 527.

\bibitem{bloedsinn} G. Barenboim, J. Bernabeu and M. Raidal, Nucl.\ Phys.\
  {\bf B511} (1998) 577.

\bibitem{JMF} J.-M.\ Fr\`{e}re et al., Phys.\ Rev.\ D {\bf 46} (1992)
  337.

\bibitem{brancolavoura} G.C.\ Branco and L. Lavoura, Phys. Lett. B
  {\bf 165} (1985) 327.

\bibitem{Chang}
%A MINIMAL MODEL OF SPONTANEOUS CP VIOLATION WITH THE GAUGE GROUP SU(2)-L X
%SU(2)-R X U(1)-(B-L).
D. Chang, Nucl. Phys. {\bf B214} (1983) 435.

\bibitem{EGZPC} G. Ecker and W. Grimus, Z.\ Phys.\ C {\bf 30} (1986)
  293.

\bibitem{sinead} S. Ryan, Talk given at {\em 1999 Chicago Conference
    on Kaon Physics (K 99)}, Chicago (IL),  June 1999, 
Preprint hep--ph/9908386.

\bibitem{msSRs} M. Jamin and M. M\"unz, Z. Phys.\ C {\bf 66} (1995)
  633;\\
K.G.\ Chetyrkin, D. Pirjol and K. Schilcher, Phys.\ Lett.\ B {\bf 404}
  (1997) 337;\\
P. Colangelo et al., Phys.\ Lett.\ B {\bf 408} (1997) 340;\\
M. Jamin, Nucl.\ Phys.\ B Proc.\ Suppl.\ {\bf 64} (1998) 250.

\bibitem{PDG} C. Caso et al. (PDG), Eur.\ Phys.\ J.\ C {\bf 3} (1998) 1.

\bibitem{higgsbound} F.I.\ Olness and M.E.\ Ebel, Phys.\ Rev.\ D {\bf
    32} (1985) 1769.

\bibitem{GL} M. Gronau and D. London, Phys.\ Rev.\ D {\bf 55} (1997) 2845.

\bibitem{Misiak}
%B ---> S GAMMA DECAY IN SU(2)-L X SU(2)-R X U(1) EXTENSIONS OF THE STANDARD
%MODEL.
P. Cho and M. Misiak, Phys. Rev. D {\bf 49} (1994) 5894.

\bibitem{natural} G. Ecker, W. Grimus and W. Konetschny, Phys.\ Lett.\
  B {\bf 94} (1980) 381; Nucl.\ Phys.\ {\bf B177} (1981) 489.

\bibitem{xx} G. Barenboim et al., Phys.\ Rev.\ D {\bf 60} (1999) 016003.

\bibitem{JMF2} J.-M.\ Fr\`{e}re et al., Phys.\ Rev.\ D {\bf 45} (1992)
  259. 

\bibitem{df2} J. Bijnens, J.-M.\ G\'{e}rard and G. Klein, Phys.\ Lett.\ B 
{\bf 257} (1991) 191.

\bibitem{babar} The {\sc BaBar} Physics Book, P.F.\ Harrison and H.R.\
  Quinn (eds.), SLAC Report 504 (1998).

\bibitem{buhmann} G. Buchalla, A. Buras and M. Lautenbacher, Rev.\
  Mod.\ Phys.\ {\bf 68} (1996) 1125.

\bibitem{laurent} L. Lellouch, Talk given at {\em 34th Rencontres de
    Moriond}, Les Arcs, France, March 1999, Preprint
CERN--TH/99--140 (hep--ph/9906497).

\bibitem{soni} W.-S. Hou and A. Soni, Phys.\ Rev.\ D {\bf 32} (1985)
  163.

\bibitem{gauge}
J. Basecq, L.-F.\ Li and P.B.\ Pal, Phys.\ Rev.\ D {\bf 32} (1985)
  175.

\bibitem{V} A.I.  Vainshtein et al., Sov.\ J.\ Nucl.\ Phys.\ {\bf 23}
  (1976) 540 [Yad.\ Fiz.\ {\bf 23} (1976) 1024];\\
M.I.\ Vysotsky, Sov.\ J. Nucl.\ Phys.\ {\bf 31} (1980)
  797 [Yad.\ Fiz.\ {\bf 31} (1980) 1535].

\bibitem{paschos} A. Datta et al., Preprint hep--ph/9509420 (unpublished).

\bibitem{osc} The LEP B Oscillation Working Group, Preprint LEPBOSC 98/3.

\bibitem{vancouver} J. Alexander, Plenary Talk given at ICHEP 98,
  Vancouver, July 1998,\\ 
{\tt http://ichep98.triumf.ca/private/convenors/transparencies/plenary7.pdf}.

\bibitem{CDF} M.P.\ Schmidt (CDF), Talk given at {\em 34th Rencontres
      de Moriond}, Les Arcs, France, March 1999, Preprint
      Fermilab--Conf--99/157--E (hep--ex/9906029).

\bibitem{nakada} T. Nakada, Preprint PSI--PR--91--02 (unpublished).

%\bibitem{B4}
%G.~Barenboim, J.~Bernabeu and M.~Raidal,
%``CP asymmetries in B0 decays in the left-right model,''
%Phys. Rev. Lett. {\bf 80} (1998) 4625.

\bibitem{bardeen} W.A.\ Bardeen, Proceedings of the {\em International 
Symposium on Heavy Flavor and Electroweak Theory}, Beijing, P.R.\
China,  August 1995, p.\ 88 (Preprint Fermilab--Conf--95/378--T).

\bibitem{SVZ} M.A.\ Shifman, A.I.\ Vainshtein and V.I.\ Zakharov,
Nucl.\ Phys.\ {\bf B147} (1979) 385; 448; 519.

\bibitem{QCDBB} 
A.A.\ Ovchinnikov and A.A.\ Pivovarov, Sov.\ J.\ Nucl.\ Phys.\  {\bf
  48} (1988) 120 [Yad.\ Fiz.\ {\bf 48} (1988) 189]; Phys.\ Lett.\ B {\bf
  207} (1988) 333;\\
S. Narison and A.A.\ Pivovarov, Phys.\ Lett.\ B {\bf 327} (1994) 341;\\
A.A. Pivovarov, Talk given at the {\em 3rd German-Russian Workshop on
  Progress in Heavy Quark Physics}, Dubna, Russia, May 1996, Preprint 
hep--ph/9606482. 

\bibitem{bestseller} E. Bagan et al., Phys.\ Lett.\  B {\bf 278}
  (1992) 457.

%\bibitem{abel} S.A.~Abel and J.-M.~Fr\`{e}re,
%Phys. Rev. D {\bf 55} (1997) 1623 (hep--ph/9608251).

\end{thebibliography}
\end{document}